\documentclass[twoside,11pt]{article}

\usepackage[abbrvbib, preprint]{jmlr2e}

\usepackage{cite}
\usepackage{amsmath,amssymb,amsfonts}

\usepackage{algorithmic}
\usepackage{sgame}

\usepackage{graphicx}
\usepackage{textcomp}
\usepackage{xcolor}
\def\BibTeX{{\rm B\kern-.05em{\sc i\kern-.025em b}\kern-.08em
		T\kern-.1667em\lower.7ex\hbox{E}\kern-.125emX}}
\usepackage{float}

\newtheorem{assumption}{Assumption}[section]

\usepackage{caption} 
\usepackage{subcaption} 
\usepackage{tikz}

\usepackage{amssymb,latexsym,amsmath}
\usepackage{epsfig}
\usepackage{graphicx}%
\usepackage{color}%
\usepackage{dsfont}%
\usepackage{bbold}

\usepackage{lipsum}
\usepackage{graphicx}
\usepackage{epstopdf}
\ifpdf
\DeclareGraphicsExtensions{.eps,.pdf,.png,.jpg}
\else
\DeclareGraphicsExtensions{.eps}
\fi

\usepackage{fancyhdr,amsmath,times,amsxtra,amssymb}
\usepackage{mathrsfs}

\usepackage{amstext}
\usepackage{amssymb,amsfonts}
\usepackage{mathtools}

\usepackage{graphicx} 
\usepackage{epstopdf}
\usepackage{verbatim}

\usepackage{color}

\usepackage{fancybox}

\DeclareGraphicsExtensions{.eps}
\usepackage{tikz}
\usetikzlibrary{arrows,positioning}

\allowdisplaybreaks

\usepackage{sgame}
\usepackage{upgreek}
\usepackage{bm}
\usepackage[linesnumbered, ruled,noend]{algorithm2e}

\usepackage{graphicx}

\usepackage{mathrsfs}  

\newcommand{\nn}{\mathbb{N}}
\newcommand{\rr}{\mathbb{R}}
\newcommand{\uu}{\mathbb{U}}
\newcommand{\xx}{\mathbb{X}}

\newcommand{\DD}{\mathcal{D}}

\newcommand{\GG}{\mathcal{G}}

\newcommand{\NN}{\mathcal{N}}
\newcommand{\PP}{\mathcal{P}}

\newcommand{\bU}{\textbf{U}}
\newcommand{\bu}{\textbf{u}}
\newcommand{\ba}{\textbf{a}}

\newcommand{\bs}{\textbf{s}}
\newcommand{\bT}{\textbf{T}}

\fussy

\renewcommand{\aa}{\mathbb{A}}
\newcommand{\pp}{\mathbb{P}}

\newcommand{\CC}{\mathcal{C}}
\newcommand{\EE}{\mathcal{E}}

\newcommand{\best}{{\rm Best}}
\newcommand{\better}{{\rm Better}}

\newcommand{\Sat}{{\rm Sat}}

\newcommand{\Uniform}{{\rm Uniform}}
\newcommand{\UnSat}{{\rm UnSat}}

\usepackage{bm} 

\newcommand{\br}{{\mathbf{r}}}

\newcommand{\bA}{{\mathbf{A}}}

\usepackage{xcolor}

\usepackage{lastpage}

\allowdisplaybreaks

\ShortHeadings{Satisficing Paths to Equilibrium}{Yongacoglu, Hickey, Arslan, Pavel and Y\"uksel}
\firstpageno{1}

\begin{document}
	\sloppy
	\title{Satisficing Paths to Equilibrium, Generalized Weakly Acyclic Games, and Learning}
	
	\author{Bora Yongacoglu, Gwendolen Hickey, G\"urdal Arslan, Lacra Pavel, and Serdar Y\"uksel
		\thanks{B. Yongacoglu was with the Department of Electrical and Computer Engineering at the University of Toronto. G. Hickey is with the Department of Mathematics and Statistics, Queen's University. L. Pavel is with the Department of Electrical and Computer Engineering at the University of Toronto. G. Arslan is with the Department of Electrical Engineering, University of Hawaii at Manoa. S. Y\"uksel is with the Department of Mathematics and Statistics, Queen's University. Correspondence to {\tt gurdal@hawaii.edu}.}%
	}

	\maketitle
	
	\begin{abstract}
		Weakly acyclic games generalize potential games and have shown to be fundamental in the study of multi-agent learning as they allow for convergence to an equilibrium via best-responding under inertia. In this paper, we present a generalization of weakly acyclic games, and we demonstrate its importance in multi-agent learning when agents employ experimental strategy updates in periods where they fail to best respond. While weak acyclicity is defined in terms of path connectivity properties of a game's better response graph, our concept is defined using a generalized better response graph under revision dynamics termed as {\it satisficing}. We refer to this class of games as {\it generalized weakly acyclic games} (GenWAGs). We provide sufficient conditions for this notion of generalized weak acyclicity in both two-player games and $n$-player games in normal form, including static and dynamic games. Several graph theoretic characterizations of such games are presented together with sufficiency conditions, examples, and counterexamples. Finally, implications on learning via policy revision processes are presented. 
	\end{abstract}
	

	\section{Introduction}
	
	This paper presents a new class of games relevant to game theoretic learning algorithms. We refer to this class of games as {\it generalized weakly acyclic games} (GenWAGs), since they constitute a meaningful generalization of weakly acyclic games and are defined in an analogous manner \citet{young1993evolution,milchtaich1996congestion,young1998individual,young2004strategic,fabrikant2013structure}. In turn, weakly acyclic games generalize potential games \citet{monderer1996potential}, a class of games used to model cooperative and distributed control \citet{marden2009cooperative,li2013designing,gopalakrishnan2014potential}.
	
	Whereas weakly acyclic games are defined in terms graph theoretic properties of a game's better response graph, GenWAGs are defined using a game's \emph{satisficing graph}, introduced in this paper, which contains the game's better response graph as a subgraph. The definition we propose for a game's satisficing graph is based on the concept of \emph{satisficing paths}, first presented in the context of multi-state Markov games in \citet{yongacoglu2023satisficing}. In that work, a satisficing path is any sequence of strategy profiles for which the strategy of an optimizing agent (that is, an agent whose strategy is a best response to that of its counterparts at a given time) is not altered in the next period. Despite  thematic similarities, there are salient differences between the graphs studied here and the paths studied in \citet{yongacoglu2023satisficing}. In the latter, paths are defined on the set of randomized/mixed strategies, and continuity arguments play an important role in the analysis. By contrast, satisficing graphs are defined here on the set of pure strategies, our analysis centres on discrete objects, and our arguments do not hinge on continuity. Moreover, while \citet{yongacoglu2023satisficing} provided some sufficient conditions for existence of satisficing paths to equilibrium, necessary conditions were not provided and it was left open whether pathological counterexamples exist. 
	
	There are several factors motivating the study of generalized weakly acyclic games. \citet{ArslanYukselTAC16} showed that for weakly acyclic games, a graph theoretic study of policy revision processes leads to a decentralized Q-learning implementation (with only local action information) that ensures convergence to equilibria. As we show in Theorem~\ref{theorem:Markov-chain}, GenWAGs arise naturally in the analysis of certain learning-relevant stochastic processes on the set of strategy profiles in a game, with important special cases such as randomized variants of inertial better/best response dynamics. A key insight of this result is that Nash convergence can be guaranteed in a wider class of games by incorporating experimental (possibly suboptimal) strategy revision when failing to best respond, rather than rigidly requiring players to revise their strategies to better or best responses. A second motivation for studying generalized weakly acyclic games is the relative simplicity of verifying sufficient conditions: for instance, verifying the existence of a strict pure Nash equilibrium in a two-player game (Lemma \ref{theorem:two-player-sufficient-condition}) or symmetry conditions in an $n$-player game (\citet[Theorem 3.6]{yongacoglu2023satisficing}) is often more practical than verifying path connectivity properties of the game's better response graph. 
	
	In view of the decentralized Q-learning implementation of \citet{ArslanYukselTAC16}, it is worth noting that standard Q-learning has been applied to several specific multi agent applications \citet{Tan,Sen}. The long-term behavior for stateless repeated games, for which Q-learning may reduce to averaging dynamics, has been analyzed in \citet{Leslie03b}, and is strongly connected to the long-term behavior of the Stochastic Fictitious Play (SFP) dynamics \citet{fudlev} in the case of two decision makers (DMs); see Lemma~4.1 in \citet{Leslie03b}. 
	Using the connection between Q-learning dynamics and SFP dynamics, the convergence of Q-learning has been established in zero-sum games as well as in team games with two DMs; see Proposition~4.2 in \citet{Leslie03b}.
	%
	Fictitious Play dynamics \citet{bro,rob} have been extended to stochastic games in several other works, including \citet{Claus,Schoenmakers,fp-stochastic-games} and more recent works such as \citet{swenson2018distributed, eksin2017distributed}. While these results provide further insight into learning dynamics in stochastic games, most prior work on Fictitious Play has focused on settings with perfect monitoring of other players' actions, in contrast to the general setting we consider here.

	Due in part to the challenges posed by non-stationarity and decentralized information, most contributions to the literature on independent learners have focused either on the stateless case of repeated games and produced formal results, such as in the works of \citet{leslie2005individual, foster2006regret, germano2007global, chasparis2013aspiration, marden2012revisiting,  marden2014achieving}, or otherwise studied the multi-state setting and presented only empirical results.
	For example, \citet{daskalakis2021independent} studies the convergence of single-agent policy gradient algorithms for episodic two-player zero-sum games, and \citet{sayin2021decentralized} peforms a complementary analysis via an ODE method \citet{bormey00a} under decentralized information with two-time scales (\citet{borkar1997stochastic,borkar2002reinforcement,leslie2003convergent,leslie2005individual}). The case of non-zero-sum games was also considered recently in \citet{mao2023provably}, and further recent studies for a variety of classes of game, including team games, include \citet{baudin2022fictitious,zhang2020model,zhang2021multi,zhang2018networked,mao2023provably,sha2022fully}. Despite these advances, establishing convergence guarantees for decentralized learning dynamics in more general stochastic games remains challenging, motivating the approach considered in this work. We also note that algorithms for convergence to Nash equilibrium play an important role in the game theoretic approach to distributed control \citet{frihauf2011nash,salehisadaghiani2016distributed,ye2017distributed,gadjov2018passivity} as well.\\

	\noindent \textbf{Contributions.} 
	\begin{itemize}
		\item[(i)] This paper introduces a class of games which allow for convergence to equilibria via a satisficing policy revision process. We show that this class generalizes weakly acyclic games (which include team games and potential games); in view of this we refer to these games as Generalized Weakly Acyclic Games (GenWAGs). 
		\item[(ii)] In Theorem~\ref{theorem:Markov-chain} we show that GenWAGs coincide exactly with the class of games for which a specific Markov chain converges to a pure Nash equilibrium. 
		\item[(iii)] In Theorem \ref{theorem:2-player-sufficient-condition} for 2-players, and Theorem \ref{Cond0Thm} for $n$-players, we provide sufficient conditions for guaranteeing that a normal-form game is a GenWAG. We also provide the first negative results in the theory of satisficing and demonstrate that our generalization is non-trivial: we provide an example of a game that is not weakly acyclic but is a GenWAG, showing that the generalization is strict, and we provide examples of games that are not GenWAGs but nevertheless admit a pure Nash equilibrium, showing that our class of GenWAGs is a strict subset of the set of games admitting a pure Nash equilibrium. We also show that the normal game framework does carry over to the Markov game setting by viewing stationary policies as actions, and we show that GenWAGs strictly generalize weakly acyclic games.
		\item[(iv)] We present positive implications for multi-agent learning with local action/policy information: In particular, for stochastic games with local information under stationary policies, we show that a Q-learning policy revision process similar to that shown in \citet{ArslanYukselTAC16} converges to equilibrium if the game is GenWAG. We further discuss extensions to the case with mixed strategies, $\epsilon$-satisficing, and present several research directions. 
		\item[(v)] We thus provide a systematic characterization of games which allow for convergence to equilibria, and an associated stochastic analysis ensuring independent learning (where there is no action information sharing) towards equilibria. 
	\end{itemize}

	\section{Normal-Form Games}  \label{sec:model}

	Our setting is that of finite $n$-player normal-form games. An $n$-player game $\Gamma$ is described by a triple
	\[
	\Gamma = \left( n, \bA, \br \right) ,
	\]
	where $n$ is the number of players, $\bA = \aa^1 \times \cdots \times \aa^n$ is a finite set of action profiles (in a deterministic game this is typically be given with action profiles, whereas in stochastic games a properly defined policy profile depending on the information structure is what determines equilibria or optimality), and $\br = \{ r^i \}_{i =1}^n$ is a collection of reward functions, with $r^i : \bA \to \rr$ being player $i$'s reward function. The $i^{\rm th}$ component of $\bA$ is player $i$'s set of actions/pure strategies $\aa^i$.
	
	\ 
	
	\noindent\textbf{Notation.} We use $[n] := \{ 1, \dots, n\}$ to denote the set of players. For an element $\ba \in \bA$, we write $\ba = ( a^i )_{i \in[n]}$. To isolate the role of player $i$ we write $\ba = (a^i, \ba^{-i})$, so that $\ba^{-i}$ is interpreted as $(a^1, \dots, a^{i-1}, a^{i+1}, \dots, a^n)$. In a slight abuse of notation, we write $\bA = \aa^i \times \bA^{-i}$. For a given player $i\in[n]$, we refer to the remaining players in $[n]\setminus\{i\}$ as $i$'s counterplayers or counterparts.

	\ 
	
	\noindent\textbf{Description of play.} Each player $i \in [n]$ selects its own action $a^i \in \aa^i$, resulting in an action profile $\ba = (a^i)_{i=1}^n$. Once this action profile has been selected, each player $i \in [n]$ receives a reward $r^i ( \ba ) = r^i ( a^i, \ba^{-i})$. Player $i$'s objective is to maximize its reward $ r^i ( a^i, \ba^{-i})$ by optimizing over its action choice $a^i \in \aa^i$. Since player $i$'s objective function depends on the action selections of its counterplayers, we have the following definitions of better and best responding.
	
	\begin{definition}
		For player $i \in [n]$ and an action profile $(a^i  , \ba^{-i}) \in \bA$, an action $a^i_{\star} \in \aa^i$ is called a \emph{better response to} $(a^i  , \ba^{-i}) $ if $ r^i ( a^i_{\star} , \ba^{-i} ) \geq r^i ( a^i, \ba^{-i} ). $
		
		\vspace{5pt}
		
		\noindent If $r^i ( a^i_{\star}, \ba^{-i} ) \geq r^i ( \bar{a}^i, \ba^{-i} )$ for any $\bar{a}^i \in \aa^i$, then the action $a^i_{\star}$ is called a \emph{best response to} $\ba^{-i}$.         
	\end{definition}

	We let $\better^i (\ba) \subseteq \aa^i$ denote the subset of player $i$'s pure actions that are better responses to $\ba = (a^i, \ba^{-i})$, and we let $\best^i (\ba^{-i}) \subseteq \aa^i$ denote the subset of player $i$'s pure actions that are best responses to $\ba^{-i}$.

	For an action profile $\ba$ and a player $i \in [n]$, we say that player $i$ is \emph{satisfied} at $\ba$ if $a^i \in \best^i(\ba^{-i})$, and otherwise we say that player $i$ is unsatisfied at $\ba$. We let $\Sat(\ba) \subseteq [n]$ denote the subset of players who are satisfied at an action profile $\ba$, and let $\UnSat(\ba) \subseteq [n]$ denote the subset of players who are unsatisfied at $\ba$. 
	
	\begin{definition}
		An action profile $\ba_{\star} \in \bA$ is called a \emph{(pure) Nash equilibrium} if $a^i_{\star} \in \best( \ba^{-i}_{\star})$ for all $\forall i \in [n] . $ For a game $\Gamma$, we let ${\rm Nash}(\Gamma) $ denote the set of pure Nash equilibria of $\Gamma$.
	\end{definition}
	

	\section{Graph Theoretic Structure in Games} \label{sec:graph-theoretic-structure}
	\subsection{Best Response Paths and Graphs}
	
	To formalize the concepts of better and best response paths mentioned in the introduction, we now introduce the best response graph and better response graph of the game $\Gamma$. In what follows, all graphs are directed, and our notational conventions are such: $D = ( V, E )$ represents a directed graph, where $V$ represents the finite set of vertices of $D$ and $E \subseteq V \times V$ represents a collection of directed edges, with $(v_1, v_2 ) \in E$ meaning there is a directed edge from $v_1$ to $v_2$ in $D$.

	\begin{definition}  \label{def:best-response-graph}
		The (multi-agent) \emph{best response graph} of the game $\Gamma$ is a directed graph $\DD_{\best} ( \Gamma ) = ( \bA , \EE_{\best} )$, where, for any $(\ba_1  , \ba_2 ) \in \bA \times \bA$, we have $(\ba_1, \ba_2) \in \EE_{\best}$ if and only if the following conditions hold for each player $i \in [n]$: 
		\begin{itemize}
			\setlength \itemsep{2pt}
			
			\item[1.] $a^i_1 \in \best^i ( \ba^{-i}_1 ) \Rightarrow a^i_2 = a^i_1 $, and 
			
			\item[2.] $a^i_2 \not= a^i_1 \Rightarrow a^i_2 \in \best^i (\ba^{-i}_1 )$. 
		\end{itemize}
	\end{definition}
	
	Intuitively, a directed edge from action profile $\ba_1$ to $\ba_2$ exists when $\ba_2$ is obtained by switching the strategies of players in some subset $\CC_1 \subseteq \UnSat(\ba_1)$, and when for each such player $i \in \CC_1$, $a^i_2$ belongs to $\best^i (\ba^{-i}_1 )$. 
	
	We also note that this construction allows for the actions of several players to be changed simultaneously, and further that it is acceptable that a player does not change the policy even when he is not best responding. This is needed for allowing for inertia to avoid cyclical behaviour. 
	
	Consider the $2\times2$ discoordination game in Figure~\ref{fig:example1-game}. In this game, Player 1 selects the row, Player 2 selects the column. Player 1 is paid the first quantity in the chosen cell, and Player 2 receives the second quantity. Player 1's best response is to copy the action of Player 2, and Player 2's best response is to mismatch the action of Player 1.

	\begin{figure}[h]
		\centering
		\begin{subfigure}[b]{0.3\textwidth}
			\centering
			\begin{game}{2}{2}
				& $a$       & $b$       \\
				$A$     & 1, 0      & 0, 1      \\
				$B$     & 0, 1      & 1, 0       \\
			\end{game}%
			\vspace{5pt}
			\caption{Discoordination game}  \label{fig:example1-game}
		\end{subfigure}
		\begin{subfigure}[b]{0.3\textwidth}
			\centering
			\begin{tikzpicture}
				\node[circle, draw] (ba) at (0,0) {};
				\node[circle, draw] (bb) at (1,0) {};
				\node[circle, draw] (ab) at (1,1) {};
				\node[circle, draw] (aa) at (0,1) {};
				
				\draw[->, >=stealth, line width=0.75pt] (ba) -- (aa);
				\draw[->, >=stealth, line width=0.75pt] (aa) -- (ab);
				\draw[->, >=stealth, line width=0.75pt] (ab) -- (bb);
				\draw[->, >=stealth, line width=0.75pt] (bb) -- (ba);
			\end{tikzpicture}
			\caption{Best response graph}   \label{fig:example1-graph}
		\end{subfigure}
		\caption{A discoordination game and its best response graph, with self-loops omitted.}
	\end{figure}
	
	The best response graph of this game is displayed in Figure~\ref{fig:example1-graph}, with node labels (e.g. $(A,a)$, etc.) and self-loops omitted for visual clarity. For example, there is a directed edge $(B,a) \to (A,a)$, because action $A \in \best^1 ( a )$. On the other hand, there is no directed edge from $(B,a) \to (A,b)$, because $a \in \best^2 ( B )$ and thus player 2 is satisfied at $(B,a)$.
	
	A (multi-agent) \emph{best response path} in the game $\Gamma$ is defined as any path in the directed graph $\DD_{\best} ( \Gamma)$.

	Next, we define the better response graph for the game $\Gamma$. This construction is similar to the best response graph, but allows for suboptimal policy revision when a player is not satisfied. 
	
	\begin{definition}   \label{def:better-response-graph}
		The (multi-agent) \emph{better response graph} of the game $\Gamma$ is a directed graph $\DD_{\better} ( \Gamma ) = ( \bA , \EE_{\better} )$, where $\EE_{\better} \subseteq \bA \times \bA$ is characterized as follows: for a pair $(\ba_1  , \ba_2 ) \in \bA \times \bA$, one has $(\ba_1, \ba_2) \in \EE_{\better}$ if and only if both of the following hold for each player $i \in [n]$, 
		\begin{itemize}
			\setlength \itemsep{2pt}
			
			\item[1.] $a^i_1 \in \best^i ( \ba^{-i}_1 ) \Rightarrow a^i_2 = a^i_1 $, and 
			
			\item[2.] $a^i_2 \not= a^i_1 \Rightarrow a^i_2 \in \better^i (\ba_1 )$. 
		\end{itemize}
	\end{definition}
	
	A multi-agent better response path is defined as a path in the directed graph $\DD_{\better}(\Gamma)$. We note that in the discoordination game of Figure~\ref{fig:example1-game}, the better and best response graphs coincide ($\DD_{\best}(\Gamma) = \DD_{\better}(\Gamma)$) but this is not generally the case.

	With the preceding definitions in hand, we are now ready to present the definition of weakly acyclic games, which have been defined in several related forms of differing generality \citet{young1993evolution,milchtaich1996congestion,young1998individual,young2004strategic,fabrikant2013structure}.
	
	\begin{definition} 
		A game $\Gamma$ is called \emph{weakly acyclic} if, for any action profile $\ba \in \bA$, there exists a multi-agent better response path beginning at $\ba$ and ending at a pure Nash equilibrium. 
	\end{definition}
	
	In graph theoretic terms, this definition has two main parts. First, the better response graph $\DD_{\better} ( \Gamma)$ must possess at least one sink (a node with no outgoing vertices), which corresponds to the existence of pure Nash equilibrium. Second, there must exist a directed path in $\DD_{\better} ( \Gamma)$ from any non-sink node to some sink node. 
	
	We observe that the discoordination game of Figure~\ref{fig:example1-game} is not a weakly acyclic game, since it possesses no pure Nash equilibrium and thus fails the first part of the definition. There are also examples of games that admit pure Nash equilibrium but are nevertheless not weakly acyclic because they fail the second condition on the existence of paths to pure equilibrium. For one such example, consider the game in Figure~\ref{fig:generalized-not-merely-wag}. This game admits a unique pure Nash equilibrium, $(T,L)$, but is not weakly acylic because there are no multi-agent better response paths from (for instance) the initial action profile $(M,C)$ to $(T,L)$. 
	
	\begin{figure}[h]
		\centering
		\begin{game}{3}{3}
			& $L$       & $C$       & $R$       \\
			$T$     & 9, 9      & 0, 0      & 0, 0      \\
			$M$     & 0, 0      & 2,1       & 1,2 \\
			$B$     & 0, 0      & 1,2       & 2,1 \\
		\end{game}%
		\vspace{5pt}
		\caption{A game with a pure Nash equilibrium that is not weakly acyclic}  \label{fig:generalized-not-merely-wag}
	\end{figure}
	
	Weakly acyclic games appear in many studies on multi-agent game theoretic learning with distributed and/or decentralized information, e.g. \citet{marden2009payoff,ArslanYukselTAC16,swenson2018distributed}. One reason for their practical relevance is that weakly acyclic games are the largest class of games for which randomized inertial better response dynamics is guaranteed to converge to Nash equilibrium. That is, suppose an initial action profile $\ba_1 \in \bA$ is selected arbitrarily, and then for each time $t \geq 1$, the strategy of every player $i \in [n]$ is set according to the following randomized update rule:
	\[
	a^i_{t+1} = %
	\begin{cases} 
		a^i_t,                      \text{ if } a^i_t \in \best^i (\ba^{-i}_t ) , \\ 
		a^i \sim \Uniform( \better^i ( \ba_t ) \cup \{ a^i_t \} ), \text{ else.}
	\end{cases}
	\]
	(We have included $a^i_t$ in the latter case as a substitute for a random inertia condition, as discussed in \citet{ArslanYukselTAC16,swenson2018distributed}.)
	
	Although randomized, distributed inertial better response dynamics converge in weakly acyclic games and have other desirable qualities, such as being individually rational, one obvious shortcoming is that such algorithms do not lead to Nash equilibrium strategies in games lacking the weakly acyclic structure. This remains true even when some strategy profiles are Pareto optimal, such as the Nash equilibrium $(T,L)$ in the game of Figure~\ref{fig:generalized-not-merely-wag}. Thus, beyond weakly acyclic games, one must rely on different graph theoretic structure when designing Nash-seeking algorithms. This deficiency of the weakly acyclic structure is addressed in the next sections.

	\subsection{Satisficing Paths and Graphs}
	
	Having established the need to identify structure beyond that of weakly acyclic games, we now present the notion of the \emph{satisficing graph} of a game. The contents of this section are thematically related to the notion of satisficing paths presented in \citet{yongacoglu2023satisficing}, but here we consider only pure strategies and define explicit graphs for the first time.

	\begin{definition}  \label{def:satisficing-graph}
		The (multi-agent) \emph{satisficing graph} of the game $\Gamma$ is a directed graph $\DD_{\Sat} ( \Gamma ) = ( \bA , \EE_{\Sat} )$, where, for any $(\ba_1  , \ba_2 ) \in \bA \times \bA$, we have $(\ba_1, \ba_2) \in \EE_{\Sat}$ if and only if the following conditions hold for each player $i \in [n]$: 
		\begin{itemize}
			\setlength \itemsep{2pt}
			
			\item[1.] $a^i_1 \in \best^i ( \ba^{-i}_1 ) \Rightarrow a^i_2 = a^i_1 $, and 
			
			\item[2.] $a^i_2 \not= a^i_1 \Rightarrow a^i_2 \in \aa^i$. 
		\end{itemize}
	\end{definition}
	
	The second condition in the definition above is redundant, but was included to elucidate the successive generalization from the best response graph (Definition~\ref{def:best-response-graph}) to the better response graph (Definition~\ref{def:better-response-graph}) and then from the better response graph to the satisficing graph (Definition~\ref{def:satisficing-graph}) of a game: since $\best^i ( \ba^{-i}_1) \subseteq \better^i ( \ba_1) \subseteq \aa^i$, we immediately have
	\begin{align}
		\EE_{\best} \subseteq \EE_{\better} \subseteq \EE_{\Sat} . \label{eq:edge-set-containment}
	\end{align}
	
	Intuitively, a directed edge from action profile $\ba_1$ to action profile $\ba_2$ exists when $\ba_2$ is obtained by switching the strategies of players in some subset $\bar{\CC}_1 \subseteq \UnSat ( \ba_1 )$, and when for each player $i \in \bar{\CC}_1$, the action $a^i_2$ can take any value. In particular, $a^i_2$ need not be a better or best response to $\ba^{-i}_2$.
	
	A \emph{(pure) satisficing path} for the game $\Gamma$ is defined as any directed path in $\DD_{\Sat}$. From \eqref{eq:edge-set-containment}, one has that any best response path (or any better response path) is also a satisficing path, but the reverse is not generally true.

	\begin{definition} 
		A game $\Gamma$ is called a \emph{generalized weakly acyclic game} (GenWAG) if, for any action profile $\ba \in \bA$, there exists a satisficing path beginning at $\ba$ and ending at a pure Nash equilibrium. 
	\end{definition}
	
	\ 
	
	We now state some simple but useful results, which show that GenWAGs are a meaningful generalization of weakly acyclic games. First, in Theorem~\ref{lemma:WAGs-are-GenWAGs}, we observe that all weakly acyclic games are generalized weakly acyclic, but the converse does not typically hold. Then, we provide an example of a game that admits pure Nash equilibrium but is not generalized weakly acyclic. 
	
	\begin{theorem} \label{lemma:WAGs-are-GenWAGs}
		If a game $\Gamma$ is weakly acyclic, then it is also generalized weakly acyclic. Moreover, there are games that are generalized weakly acyclic but not weakly acyclic.
	\end{theorem}
	
	\noindent\textbf{Proof.} The first part follows from the fact that any multi-agent better response path is automatically a pure satisficing path. For the second part, we observe that the game presented in Figure~\ref{fig:generalized-not-merely-wag} is not weakly acyclic but is indeed generalized weakly acyclic. \hfill $\diamond$
	
	\ 
	
	Since $\EE_{\Sat}$ is defined by a rather weak constraint, it is natural to ask whether \emph{all} games with pure Nash equilibrium  are generalized weakly acyclic. We now present an example to show that the class of generalized weakly acyclic games is not trivial. That is, we show that existence of pure Nash equilibrium is not sufficient for a game to be generalized weakly acyclic. Consider the game in Figure~\ref{fig:non-trivial-game}, below. 
	
	\begin{figure}[h]
		\centering
		\begin{game}{3}{3}
			& $L$       & $C$       & $R$       \\
			$T$     & 1, 1      & 0, 1      & 0, 1      \\
			$M$     & 1, 0      & 1,0       & 0,1 \\
			$B$     & 1, 0      & 0,1       & 1,0 \\
		\end{game}%
		\vspace{5pt}
		\caption{A game that is not generalized weakly acyclic}  \label{fig:non-trivial-game}
	\end{figure}
	
	In this game, the unique pure Nash equilibrium is $(T,L)$, but the game does not admit pure satisficing paths to $(T,L)$ from any of the strategy profiles $(M,C)$, $(M,R)$, $(B,C)$, or $(B,R)$.
	
	The qualitative difference here between the games of Figure~\ref{fig:generalized-not-merely-wag} and Figure~\ref{fig:non-trivial-game} has to do with indifference. In the game of Figure~\ref{fig:non-trivial-game}, beginning at one of the strategy profiles $(M,C)$, $(M,R)$, $(B,C)$, or $(B,R)$, reaching pure Nash equilibrium $(T,L)$ requires switching the actions of both players. However, in each of these action profiles, there is exactly one satisfied player and one unsatisfied player. Upon switching the action of the unsatisfied player, one of two situations arises: either (1) the satisfied player remains satisfied, and is not compelled to change its behavior, but the unsatisfied player remains unsatisfied; or (2) the unsatisfied player becomes satisfied at the new action profile while the previously satisfied player becomes unsatisfied. In any case, there are no directed edges with tail in $\bA \setminus \{(T,L)\}$ and head $(T,L)$.

	\section{GenWAGs and Satisficing Markov Chains} \label{sec:random-walks}
	In this section, we define \emph{satisficing Markov chains}  and we characterize generalized weakly acyclic games as being exactly those games for which a satisficing Markov chain eventually converges to pure Nash equilibrium.
	
	\begin{definition} \label{def:satisficing-Markov-chain}
		For a game $\Gamma$ and an action profile $\ba \in \bA$, a \emph{satisficing Markov chain} beginning at $\ba$ is a Markov chain $\{ \ba_t \}_{t =1}^{\infty}$ on $\bA$ with $\pp(\ba_1 = \ba) = 1$ and satisfying the following evolution rule for each $t \geq 1$:
		\begin{equation}
			a^i_{t+1} = %
			\begin{cases}
				a^i_t,                      &\text{if } a^i_t \in \best^i ( \ba^{-i}_t) \\
				a^i \sim \Uniform(\aa^i ),  &\text{else,}
			\end{cases}     
			\label{eq:Markov-chain-update}
		\end{equation}
		where the collection $\{a^i_{t+1}\}_{i \in [n]}$ is jointly conditionally independent given $\ba_t$. 
	\end{definition}
	
	Definition~\ref{def:satisficing-Markov-chain} is motivated by a family of closely related game theoretic learning algorithms \citet{foster2006regret, germano2007global, young2009learning, marden2009payoff, pradelski2012learning, chasparis2013aspiration, marden2014achieving, marden2017selecting, hu2019convergence, yongacoglu2022decentralized, yongacoglu2023satisficing}. In these learning algorithms, each player $i$ has a baseline action $a^i_t$ that it periodically revises to $a^i_{t+1}$.\footnote{The subscripts $t$ and $t+1$ here denote successive revision times, which usually do not correspond to successive periods of actual play of the game.} Learning phases occur between baseline action revision times, and in a learning phase player $i$ may experiment with non-baseline actions and use the resulting reward observations to evaluate the performance of its various alternative actions, $a \in \aa^i$. At the end of such a learning phase, each player $i$ revises its baseline action from $a^i_t$ to $a^i_{t+1}$. Although the particular revision mechanism may vary, it is typically characterized by a ``win stay, lose shift'' condition similar to that of \eqref{eq:Markov-chain-update}, whereby unsatisfied agents consider randomized experimental action revision and satisfied agents continue using their previous baseline action.\footnote{The definition of ``winning'' is handled differently by different algorithms. In some cases, such as \citet{marden2014achieving} or \citet{marden2017selecting}, players maintain a \emph{mood variable} to guide their revisions. In others, such as \citet{chasparis2013aspiration} and \citet{yongacoglu2022decentralized}, players set a scalar \emph{aspiration level} and winning is defined as exceeding this aspiration level. In still others, such as \citet{yongacoglu2023satisficing} or \citet{foster2006regret}, the best response condition of \eqref{eq:Markov-chain-update} is used as the winning condition.} A number of Nash-seeking algorithms and multi-agent learning algorithms from this family, including \citet{yongacoglu2023satisficing,foster2006regret,young2009learning} and \citet{gaveau2020performance}, have been studied by first analyzing the convergence properties of a Markov chain $\{ \ba_t \}_{t =1}^{\infty}$ and then using $\{ \ba_t \}_{t =1}^{\infty}$ to approximate a sequence of learned strategy iterates $\{ \widehat{\ba}_t \}_{t =1}^{\infty}$.
	
	We now review some Markov chain terminology as it relates to Definition~\ref{def:satisficing-Markov-chain}. For action profiles $\ba, \tilde{\ba} \in \bA$, $\tilde{\ba}$ is said to be \emph{accessible} from $\ba$ if there exists a positive integer $m= m(\ba, \tilde{\ba})$ such that $\pp ( \ba_{m+1} = \tilde{\ba} | \ba_1 = \ba ) > 0$. That is, the satisficing Markov chain beginning at $\ba$ transits to $\tilde{\ba}$ in finitely many steps with non-zero probability. The action profiles $\ba $ and $ \tilde{\ba}$ are said to \emph{communicate} if $\ba$ is accessible from $\tilde{\ba}$ and $\tilde{\ba}$ is accessible from $\ba$. It is easy to see that communication defined this way allows one to partition $\bA$ into equivalence classes called \emph{communicating classes}, where $\ba$ and $\tilde{\ba}$  belong to the same communicating class if and only if $\ba$ and $\tilde{\ba}$ communicate. A communicating class $\bA^{\prime} \subseteq \bA$ is called \emph{absorbing} if the probability of transiting out of $\bA^{\prime}$ is zero when the Markov chain starts in $\bA^{\prime}$. That is, $\bA^{\prime}$ is absorbing if 
	\[
	\pp \left( \bigcup_{t \geq 1} \left\{ \ba_t \notin \bA^{\prime} \right\} \middle| \ba_1 \in \bA^{\prime} \right) = 0 .
	\]
	An action profile $\ba_{\dagger} \in \bA$ is said to be absorbing if the singleton $\{ \ba_{\dagger} \}$ is an absorbing communicating class. Before presenting the main result of this section, we prove that an action profile $\ba_{\dagger}$ is absorbing for the satisficing Markov chain if and only if it is a pure Nash equilibrium of the game.
	
	\begin{lemma} \label{lemma:absorbing-equilibrium}
		For a game $\Gamma$ and an action profile $\ba_{\dagger}$, we have that $\ba_{\dagger}$ is absorbing for the satisficing Markov chain $\{ \ba_t \}_{t =1}^{\infty}$ if and only if $\ba_{\dagger}$ is a Nash equilibrium of $\Gamma$. 
	\end{lemma}
	
	\noindent \textbf{Proof.} Suppose $\ba_{\dagger} \in \bA$ is a Nash equilibrium of $\Gamma$. By the best response condition of \eqref{eq:Markov-chain-update}, one has 
	\[ 
	\pp ( a^i_2 = a^i_{\dagger} | \ba_1 = \ba_{\dagger} ) = 1, \: \forall i \in[n]. 
	\] 
	Thus, $\pp ( \ba_2 = \ba_{\dagger} | \ba_1 = \ba_{\dagger} ) = 1^n = 1$, which shows that $\{ \ba_{\dagger} \}$ is absorbing.
	
	For the reverse direction, suppose $\{ \ba_{\dagger} \}$ is absorbing. Then, by \eqref{eq:Markov-chain-update}, the best response condition holds for each player. (Otherwise there exists a player whose action is switched with non-zero probability, which would contradict the fact that $\ba_{\dagger}$ is absorbing.) Since all players are best responding, $\ba_{\dagger}$ is a Nash equilibrium. \hfill $\diamond$
	
	\ 
	
	We now present the first of our main results, which says that GenWAGs are the largest class of games for which satisficing Markov chains are guaranteed to converge to Nash equilibrium. 
	
	\begin{theorem} \label{theorem:Markov-chain}
		A game $\Gamma$ is generalized weakly acyclic if and only if for every $\ba \in \bA$, the satisficing  Markov chain beginning at $\ba$ converges to a pure Nash equilibrium almost surely. 
	\end{theorem}
	
	\noindent\textbf{Proof.} ($\Rightarrow$) To prove the forward direction, let $\ba \in \bA$ be arbitrary, suppose $\{ \ba_t \}_{t = 1}^{\infty}$ is a satisficing Markov chain beginning at $\ba$, and suppose $\Gamma$ is generalized weakly acyclic. 
	
	Since $\Gamma$ is a GenWAG, there exists a satisficing path connecting any initial point $\tilde{\ba} \in \bA$ to a Nash equilibrium. Let $T(\tilde{\ba}) \in \nn$ be the length of a shortest such path beginning at $\tilde{\ba}$, and let $\rho(\tilde{\ba}) > 0$ be the probability that the satisficing Markov chain follows such a path when initialized at $\tilde{\ba}$. Since $\bA$ is finite, we have
	\[
	\max_{\tilde{\ba} \in \bA} T(\tilde{\ba}) =: \tau < \infty ,  \text{ and } \min_{\tilde{\ba} \in \bA} \rho ( \tilde{\ba} ) =: \rho > 0 
	\]
	
	By Lemma~\ref{lemma:absorbing-equilibrium}, any pure Nash equilibrium is a sink in the satisficing graph and thus corresponds to an absorbing state for the satisficing Markov chain.
	
	From the preceding discussion, one obtains the following inequality:
	\[
	\pp \left(  \ba_{t+\tau} \notin {\rm Nash}(\Gamma) \middle| \ba_t = \tilde{\ba}     \right) \leq (1-\rho),
	\]
	for any $\tilde{\ba} \in \bA$ and any $t \geq 1$. Recursively, one then obtains
	\[
	\pp \left( \ba_{t+m\tau} \notin {\rm Nash}(\Gamma)  \middle| \ba_t = \tilde{\ba} \right) \leq (1-\rho)^m ,
	\]
	for any $t, m \geq 1$ and $\tilde{\ba} \in \bA$. Taking $m \to \infty$, one sees that $\{ \ba_t \}_{t =1}^{\infty}$ converges to some pure Nash equilibrium almost surely for any initial condition $\ba_1 = \ba$. 
	
	($\Leftarrow$) We argue the backward direction by contrapositive. That is, we will argue that if $\Gamma$ is not a GenWAG, then there exists some initial condition $\ba_1 = \ba \in \bA$ such that the satisficing Markov chain will fail to converge almost surely to a Nash equilibrium. 
	
	Indeed, if $\Gamma$ is not a GenWAG, then there is some action profile $\ba \in \bA$ for which there are no satisficing paths beginning at $\ba$ and terminating at a pure Nash equilibrium of the game $\Gamma$. This implies that for the satisficing Markov chain, the communicating class of $\ba$ contains no Nash equilibrium action profiles, and the satisficing Markov chain beginning at $\ba_1 = \ba$ will avoid Nash equilibrium action profiles at all times. \hfill $\diamond$

	\section{Sufficient Conditions} \label{sec:existence-result}
	
	We previously encountered an example of a game, shown in Figure~\ref{fig:non-trivial-game} of Section~\ref{sec:graph-theoretic-structure}, that possessed pure Nash equilibrium but was not generalized weakly acyclic. This showed that existence of pure Nash equilibrium alone is not sufficient for a game to be generalized weakly acyclic, and constitutes the first negative example in the theory on satisficing paths and graphs. In this section, we present sufficient conditions to complement that observation.
	
	\subsection{A Sufficient Condition for Two-Player Games}
	
	\begin{definition}
		For a player $i $ and an action profile $\ba^{-i}$, an action $a^i_{\star} \in \aa^i$ is a \emph{strict best response} to $\ba^{-i}$ if
		\[
		r^i ( a^i_{\star}, \ba^{-i} ) > r^i ( a^i, \ba^{-i} ) , \quad \forall a^i \not= a^i_{\star}. 
		\]
	\end{definition}

	\begin{definition}
		A pure Nash equilibrium $\ba_{\star}$ is \emph{strict} if for each $i \in [n]$, $a^i_{\star}$ is a strict best response to $\ba^{-i}_{\star}$. 
	\end{definition}

	\begin{lemma}  \label{theorem:two-player-sufficient-condition}
		Let $\Gamma$ be a two-player game and suppose $\Gamma$ admits a strict pure Nash equilibrium $\ba_{\star}$. Then, $\Gamma$ is generalized weakly acyclic. 
	\end{lemma}
	
	\noindent \textbf{Proof.} Let $\ba_1 = \ba \in \bA$ be an arbitrary initial action profile. We proceed in cases.
	
	{\it Case 1:} If $\UnSat(\ba) = \varnothing$, then $\ba$ is a pure Nash equilibrium itself and so the path $(\ba)$ is a satisficing path beginning at $\ba$ and terminating at pure Nash equilibrium.  
	
	{\it Case 2:} If $\UnSat(\ba) = \{1,2\}$, then neither player is satisfied at $\ba$, and so $(\ba, \ba^{\prime}) \in \EE_{\Sat}$ is a valid satisficing path for any $\ba^{\prime}  \in \bA$, since (vacuously) it does not change the action of any player that was previously best responding. Selecting $\ba^{\prime} = \ba_{\star} $, one obtains a satisficing path from $\ba$ to a pure Nash equilibrium of $\Gamma$. 
	
	{\it Case 3:} $\UnSat(\ba) = \{ i \}$ for some $i \in \{1,2\}$. That is, exactly one player, $i$, is unsatisfied at $\ba$, while the other player, $j\not= i$, is satisfied at $\ba$. We put $\ba_2 = ( a^i_{\star}, a^j)$ and note $(\ba_1, \ba_2 ) \in \EE_{\Sat}$.
	
	If $a^j = a^j_{\star}$, then we have constructed our path and there is nothing left to show, since $\ba_2 = \ba_{\star}$ is a pure Nash equilibrium. Otherwise, if $a^j \not= a^j_{\star}$, then since $\ba_{\star}$ is a strict Nash equilibrium, one has that $a^j \notin \best^j ( a^i_{\star} )$. Then, we put $\ba_3 = ( a^i_{\star}, a^j_{\star}) = \ba_{\star}$, and note that $(\ba_2, \ba_3 ) \in \EE_{\Sat}$, completing the proof.   \hfill $\diamond $

		%
	
	In the following, we aim to relax the strictness condition. 
	\begin{example}
		This example shows that the existence of a pure equilibrium $\ba_\star=(a^{1}_\star,a^{2}_\star)$ where for a player $i$ (either $i=1,2$), there exists $\tilde{a}^{i}$ such that $r^i(\tilde{a}^{i},a^{-i}_\star) < r^i(a^{i}_\star,a^{-i}_\star)$ is not sufficient for the game to be GenWAG. 
		
		\begin{figure}[h]
			\centering
			\begin{game}{4}{4}
				& $a$      & $b$       & $c$ 	& $d$     \\
				$A$     & 1, 1       & 0, 1        & 0, 1  	& 0, 1    \\
				$B$     & 0, 0      & 0, 1       & 0, 1  	& 0, 1    \\
				$C$     & 1, 0      & 1, 0        & 0, 1  	& 1, 0 \\
				$D$     & 1, 0      & 0, 1        & 1, 0 	& 0, 1
			\end{game}%
			\vspace{5pt}
			\caption{A game that fails to be generalized weakly acyclic,\ despite the existence of a pure equilibrium $\ba_\star$ which some player $i$ strictly prefers over $(\tilde a^i,a^{i}_\star)$, for some  action $\tilde a^i$}  \label{fig:xxx}
			
		\end{figure}
	\end{example}
	
	The above culminate into the following theorem.
	
	\begin{theorem}\label{theorem:2-player-sufficient-condition} 
		Let $\Gamma$ be a two-player game and suppose $\Gamma$ admits a pure equilibrium $\ba_\star$ that is strict for one player. That is, without loss of generality:
		\[
		r^1(a^1_\star,a^2_\star)>r^1(a^1,a^2_\star) \quad\forall a^1\neq a^1_\star
		\]
		Then, $\Gamma$ is generalized weakly acyclic.
	\end{theorem}
	
	\noindent \textbf{Proof.}
	Let $\ba_1 = \ba \in \bA$ be an arbitrary initial action profile. We proceed in cases; {\it Case 1} and {\it Case 2} are identical to those in the proof of Lemma \ref{theorem:two-player-sufficient-condition}, and are omitted.
	
	{\it Case 3.1:} $\UnSat(\ba) = \{ 2\}$. We put $\ba_2 = ( a^1,a_\star^2)$ and note $(\ba_1, \ba_2 ) \in \EE_{\Sat}$. Because $a_\star^1$ is player 1's unique best response to $a_\star^2$, they are now unsatisfied, and we set $\ba_3=(a^1_\star,a^2_\star)=\ba_\star$.
	
	{\it Case 3.2:} $\UnSat(\ba) = \{ 1\}$. If player 1 is unable to choose any strategy that makes player 2 unsatisfied, player 1 chooses any best response $\tilde{a}^1\in \best^1(a^2)$, and we set $\ba_2=(\tilde{a}^1,a^2)$, which constitutes a Nash equilibrium. If, on the other hand, player 1 has some strategy $\bar{a}^1$ that can dissatisfy player 2, let $\ba_2=(\bar{a}^1,a^2)$. This leaves us with either {\it Case 2} or {\it Case 3.1}, so a path exists.\hfill $\diamond $

	\subsection{A Sufficient Condition for $n$-Player Games}

	To state the results for $n$-player games, we require some additional notation. For a (possibly empty) player subset $\NN \subseteq [n]$, we denote a \emph{partial} action profile $\ba^{\NN} = ( a^i )_{i \in \NN}$ as the action profile $\ba$ with components for players in the subset $\NN$. We write $\ba^{-\NN}$ to denote $\ba^{[n] \setminus \NN}$, and we then have $\ba = ( \ba^{\NN}, \ba^{-\NN})$. We let $\bA^{\NN} = \times_{i \in \NN} \aa^i$. 
	
	\begin{definition}
		Let $\Gamma = ( n, \bA, \br)$ be an $n$-player game. Let $\NN \subseteq [n]$ be an $m$-player subset, with $0 \leq  m \leq n$, and let $\ba^{-\NN}$ be a partial action profile. The \emph{subgame induced by $\ba^{-\NN}$} is an $m$-player game $\Gamma_{\dagger}$,
		\[
		\Gamma_{\dagger} = \left( m , \: \bA_{\dagger} = \bA^{\NN}, \: \br_{\dagger} \right),
		\]
		with player set $\NN$, action sets $\aa^i_{\dagger} = \aa^i$ for $i \in \NN$, and the following reward functions for each $i \in \NN$:
		\[
		r^i_{\dagger} \left( \ba^{\NN} \right) = r^i \left( \ba^{\NN} , \ba^{-\NN} \right) , \quad \forall \ba^{\NN}. 
		\]
	\end{definition}

	Intuitively, the subgame induced by $\ba^{-\NN}$ is the game that one obtains when the actions of players in $[n]\setminus \NN$ are fixed at $\ba^{-\NN}$. The players of $[n]\setminus \NN$ become fixed aspects of the environment, and the remaining players in $\NN$ play a smaller game. We say that $\Gamma_{\dagger}$ is an induced subgame of $\Gamma$ if there exists a player subset $\NN$ and a partial action profile $\ba^{-\NN}$ for which $\Gamma_{\dagger}$ is the subgame induced by $\ba^{-\NN}$.

	\vspace{5pt}

	\noindent {\it Note: In the definitions above, we allow $\NN$ to be empty and we also allow $\NN = [n]$. Thus, a game $\Gamma$ is always an induced subgame of itself.}
	%
	%
	%
	%
	
	\begin{proposition}\label{theorem:n-player-sufficient-condition}
		Let $\Gamma$ be an $n$-player game. Suppose that for any induced subgame $\Gamma_{\dagger}$ of $\Gamma$, the induced subgame $\Gamma_{\dagger}$ admits a unique pure Nash equilibrium and this Nash equilibrium is strict. Then, $\Gamma$ is generalized weakly acyclic.
	\end{proposition}
	
	We will generalize this result further below. The proof of this proposition, however, is instructive, and will guide us for our more general results to follow. Therefore, a proof is included.
	
	\noindent \textbf{Proof.} For brevity, we say that a game has \emph{the induced subgame property} (ISP) if it satisfies the condition appearing in Proposition~\ref{theorem:n-player-sufficient-condition}. That is, a game has the ISP if any induced subgame admits a unique pure Nash equilibrium and this Nash equilibrium is strict. We also remark that if a given game has the induced subgame property, then any induced subgame also has the ISP.
	
	We prove the result by induction on the number of players. The base case of $m=2$ players is a special case of Lemma \ref{theorem:two-player-sufficient-condition}. Our induction hypothesis is such: for some $m \geq 2$, if $\Gamma_{\dagger}$ is an $m$-player game that has the ISP, then $\Gamma_{\dagger}$ is generalized weakly acyclic. 
	
	Now let $\Gamma$ be an $n$-player game that has the ISP, and suppose $n = m+1$. We will argue that our induction hypothesis implies that $\Gamma$ is generalized weakly acyclic.
	
	Let $\ba_1 = \ba \in \bA$ be an arbitrary initial action profile for the game $\Gamma$. We prove the result by showing that there exists a satisficing path from $\ba_1$ to the unique Nash equilibrium of $\Gamma$, which is strict and which we denote by $\ba_{\star} \in \bA$.
	
	We begin by ruling out trivial cases. If $\UnSat(\ba_1) = [n]$, then for any $\ba^{\prime}$ we have that $(\ba_1, \ba^{\prime})$ is a valid edge in $\EE_{\Sat} ( \Gamma)$. We may then take $\ba^{\prime} = \ba_{\star}$ to be the unique Nash equilibrium of $\Gamma$, and the proof is complete in this case. Thus, we focus on the case where $\Sat(\ba_1) \not= \varnothing$, and there exists some player $i$ who is satisfied at $\ba_1$: $a^i_1 \in \best^i ( \ba^{-i}_1)$.
	
	Recalling that $m=n-1$, consider the $(n-1)$-player subgame $\Gamma_{\dagger}$ induced by fixing the action of player $i$ to be $a^i_1$. Since $\Gamma$ has the ISP and $\Gamma_{\dagger}$ is an induced subgame of $\Gamma$, it follows that $\Gamma_{\dagger}$ also has the ISP and, by the induction hypothesis, that $\Gamma_{\dagger}$ is generalized weakly acyclic. Thus, for any $(n-1)$-player strategy profile $\tilde{\ba}^{-i}_1$ of the game $\Gamma_{\dagger}$, there exists a directed path in the graph $\DD_{\Sat} ( \Gamma_{\dagger} )$ from $\tilde{\ba}^{-i}_1$ to the unique Nash equilibrium of the $(n-1)$-player game $\Gamma_{\dagger} $, denoted by $\tilde{\ba}^{-i}_k$. Denote such a path by $\tilde{\ba}^{-i}_1, \cdots, \tilde{\ba}^{-i}_k$, where each $\tilde{\ba}^{-i}_t \in \bA^{-i}$, and note that $\tilde{\ba}^{-i}_k$ is a strict Nash equilibrium for the subgame $\Gamma_{\dagger}$.
	
	Using the $(n-1)$-player joint actions $\tilde{\ba}^{-i}_1, \dots, \tilde{\ba}^{-i}_k$, we construct $n$-player strategy profiles for $\Gamma$ by fixing player $i$'s component at $a^i_1$:
	\[
	\ba_t = ( a^i_1, \tilde{\ba}^{-i}_t ), \quad \forall \, t \leq k . 
	\]
	Since player $i$'s action is fixed at $a^i_t = a^i_1$ for $t \leq k$, the sequence $(\ba_t)_{t = 1}^k$ is a valid satisficing path of the original game $\Gamma$ (i.e., in the directed graph $\DD_{\Sat} ( \Gamma )$). Depending on whether or not $\ba_k$ is a Nash equilibrium for the game $\Gamma$, we again proceed in cases to complete the proof.
	
	Since $\tilde{\ba}^{-i}_k$ was selected to be the Nash equilibrium of the induced subgame $\Gamma_{\dagger}$, for each player $j \not= i$, it holds that $a^j_k \in \best^j ( \ba^{-j}_k )$. That is, player $j$ is satisfied at $\ba^{-i}_k$. If player $i$ is also satisfied at $\ba_k$, then $\ba_k=\ba_\star$ and the proof is complete.
	
	If, on the other hand, player $i$ is not satisfied at $\ba_k$ (that is, $a^i_1 \notin \best^i ( \ba^{-i}_k )$), then we 
	define 
	\[
	\ba_{k+1} := ( a^i_{\star}, \ba^{-i}_k ).
	\]
	In other words, we change the action of player $i$ from $a^i_1$ to its component of $\ba_\star$, the unique Nash equilibrium of $\Gamma$. 
	
	Let $\Gamma^{\star}_{\dagger}$ denote the $(n-1)$-player subgame of $\Gamma$ induced by fixing player $i$'s action at $a^i_{\star}$. Since $\Gamma$ has the ISP, $\Gamma^{\star}_{\dagger}$ admits a unique pure Nash equilibrium, which is readily verified as being $\ba^{-i}_{\star} = ( a^j_{\star} )_{j \not= i}$, the action profile in which players $j\not=i$ play their components of $\ba_{\star}$, the unique Nash equilibrium of $\Gamma$. Moreover, by our induction hypothesis, $\Gamma^{\star}_{\dagger}$  is  generalized weakly acyclic. Thus, from any initial action profile $\ddot{\ba}^{-i}_1$ in the $(n-1)$-player game $\Gamma^{\star}_{\dagger}$, there exists a satisficing path  $\ddot{\ba}^{-i}_1, \dots, \ddot{\ba}^{-i}_{L}$  in the directed graph $\DD_{\Sat} ( \Gamma^{\star}_{\dagger} )$ beginning at $\ddot{\ba}^{-i}_1 $ and terminating at $\ddot{\ba}^{-i}_L := \ba^{-i}_{\star}$, the unique pure Nash equilibrium of $\Gamma^{\star}_{\dagger}$.
	
	Using $\ddot{\ba}^{-i}_1 := \ba^{-i}_{k+1} = \ba^{-i}_k$ in the discussion above, we define $n$-player strategy profiles $( \ba_{k+t} )_{t = 1}^L$ as
	\[
	\ba_{k+t} = \left( a^i_{\star}, \ddot{\ba}^{-i}_t \right) , \quad t = 1, 2, \dots, L .
	\]
	
	Since the component of player $i$ is fixed at $a^i_{k+t} = a^i_{k+1} = a^i_{\star}$, the sequence $(\ba_{k+t} )_{t = 1}^L$ is a satisficing  path of the original game $\Gamma$. Furthermore, the extended sequence \[( \ba_1, \dots, \ba_k, \ba_{k+1}, \dots, \ba_{k+L} )\] is also a satisficing  path of the game $\Gamma$. This concludes the proof, since $\ba_{\star} = \ba_{k+L}$ is the unique pure Nash equilibrium of $\Gamma$. \hfill $\diamond $
	
	
	{\it Remark.} In the preceding sections, we encountered an example of a two-player game (Figure~\ref{fig:non-trivial-game}) in which indifference posed a problem for generalized weak acyclicity. In that example, a (unique) non-strict pure Nash equilibrium existed, but this equilibrium was not accessible from all starting action profiles because players had non-unique best responses and could not be compelled to switch actions. While indifference may threaten generalized weak acyclicity, we also encountered sufficient conditions for generalized weak acyclicity based on non-indifference. The strictness hypotheses of Lemma~\ref{theorem:two-player-sufficient-condition} and Proposition \ref{theorem:n-player-sufficient-condition} assume the existence of equilibrium action profiles in which players are not indifferent between alternative actions but instead strictly prefer one alternative to the rest. In the case of two-player games, existence of a strict pure Nash equilibrium was sufficient for generalized weak acyclicity. Of note, uniqueness of Nash equilibrium was not assumed in the two-player case of Lemma \ref{theorem:two-player-sufficient-condition}. On the other hand, the preliminary multi-player result, Proposition~\ref{theorem:n-player-sufficient-condition}, assumes the existence and uniqueness of a pure Nash equilibrium and further that such an equilibrium is strict.

	It is then natural to ask whether this uniqueness assumption can be relaxed while still preserving generalized weak acyclicity, or whether a pathological counterexample exists. As it turns out, the existence of a strict pure Nash equilibrium in each induced subgame is not sufficient on its own for a game to be generalized weakly acyclic. Consider the $3$-player game shown in Figure~\ref{fig:non-acyclic-game} below, where player~3 chooses between \(X\), \(Y\) and \(Z\).
	
	\begin{figure}[h]
		\centering
		\begin{subfigure}[b]{0.3\textwidth}
			\centering
			{\small
				\begin{game}{3}{3}[X]
					& $L$       & $C$      & $R$       \\
					$T$     & 0, 0, 0   & 0, 0, 0  & 1, 1, 0   \\
					$M$     & 0, 0, 0   & 1, 1, 0  & 0, 0, 0   \\
					$B$     & 9, 9, 9   & 0, 0, 0  & 0, 0, 0   \\
				\end{game}
			}
		\end{subfigure}
		\begin{subfigure}[b]{0.3\textwidth}
			\centering
			{\small
				\begin{game}{3}{3}[Y]
					& $L$       & $C$      & $R$       \\
					$T$     & 0, 0, 0   & 0, 0, 0  & 1, 1, 0   \\
					$M$     & 1, 0, 1   & 1, 0, 1  & 0, 1, 1   \\
					$B$     & 0, 0, 0   & 0, 0, 0  & 0, 1, 1   \\
				\end{game}
			}
		\end{subfigure}
		\begin{subfigure}[b]{0.3\textwidth}
			\centering
			{\small
				\begin{game}{3}{3}[Z]
					& $L$       & $C$      & $R$       \\
					$T$     & 1, 0, 1   & 0, 1, 1  & 1, 0, 1   \\
					$M$     & 0, 0, 0   & 1, 1, 0  & 0, 0, 0   \\
					$B$     & 0, 0, 0   & 0, 1, 1  & 0, 0, 0   \\
				\end{game}
			}
		\end{subfigure}
		\caption{A game that admits a strict pure Nash equilibrium for each induced subgame which is not generalized weakly acyclic} \label{fig:non-acyclic-game}
	\end{figure}
	
	One can see that this game admits a strict pure Nash equilibrium in every induced subgame, and furthermore, all subgame equilibria for this game are strict. The unique pure Nash equilibrium for this game is $(B,L,X)$, but there exists no pure satisficing path to $(B,L,X)$ starting at any action profile where exactly two players are satisfied. This inability to reach equilibrium is possible due to the subgame equilibria not being unique in the two-player subgames induced by fixing a single player at their component of $(B,L,X)$. For example, take an initial action profile for this game to be $(M,C,X)$. Only player 3 is unsatisfied, so there are edges in the satisficing graph to $(M,C,Y)$ and $(M,C,Z)$. Player 1 and 2 remain satisfied if player 3 chooses $Z$; choosing $Y$, player 3 becomes satisfied and player 2 becomes unsatisfied. From here, player 2 is faced with a similar choice: choosing $L$, they remain unsatisfied, and choosing $R$, they become satisfied and player 1 becomes unsatisfied. In fact, for every action profile in $\Gamma$ where two players are satisfied, the unsatisfied player can only satisfy themself by making exactly one other player unsatisfied. This demonstrates the importance of the uniqueness assumption in Proposition \ref{theorem:n-player-sufficient-condition} and that it cannot be removed completely. Despite the inability to entirely remove the uniqueness assumption, it can be relaxed to yield a more general sufficient condition, as we present in the following, which generalize Proposition \ref{theorem:n-player-sufficient-condition}.
	
	As earlier, for a subset $\mathcal{N}\subset[n]$  and any action profile $\textbf{a}^{-\mathcal{N}}\in\textbf{A}^{-\mathcal{N}}$ of the players in $[n]\backslash\mathcal{N}$, the induced game $\Gamma_{\dagger}(\textbf{a}^{-\mathcal{N}})$ is obtained from $\Gamma$ by fixing the strategies of the players in $[n]\backslash\mathcal{N}$ at $\textbf{a}^{-\mathcal{N}}$.
	We consider an induced subgame $\Gamma_{\dagger}(\textbf{a}^{-\mathcal{N}})$ only for a non-empty subset $\mathcal{N}\subset[n]$; moreover, we let $\Gamma_{\dagger}(\textbf{a}^{-\mathcal{N}})=\Gamma$ when $\mathcal{N}=[n]$.  $\textrm{Nash}(\Gamma_{\dagger}(\textbf{a}^{-\mathcal{N}}))$ denotes the set of pure Nash equilibria in  $\Gamma_{\dagger}(\textbf{a}^{-\mathcal{N}})$.
	When $\mathcal{N}=\{i\}$ is a single-player subset of $[n$], $\textrm{Nash}\big(\Gamma_{\dagger}(\textbf{a}^{-\mathcal{N}})\big)=\best^i\big(\textbf{a}^{-i}\big)$.
	
	Consider the following sufficient conditions:
	
	\textbf{Condition 0:}
	$\forall \textbf{a}_1\in\textbf{A}$,  there exists some $\bar{\ba}\in\textrm{Nash}(\Gamma)$, such that for all $\mathcal{N}\subset\textrm{Sat}(\textbf{a}_1)$, if $\textbf{a}^{\mathcal{N}} \in \textrm{Nash}\big(\Gamma_{\dagger} (\bar\ba^{-\mathcal{N}})\big)
	$ then $ (\textbf{a}^{\mathcal{N}} ,\bar{\textbf{a}}^{-\mathcal{N}}) \in \textrm{Nash}(\Gamma)$. That is, if $\ba^\mathcal{N}$ is a Nash equilibrium of the induced subgame $\Gamma_{\dagger} (\bar\ba^{-\mathcal{N}})$, then $(\ba^{\mathcal{N}} ,\bar\ba^{-\mathcal{N}})$ constitutes a Nash equilibrium of $\Gamma$.
	
	\textbf{Condition 1:}
	$\exists\ba_\star\in\textrm{Nash}(\Gamma)$, such that for all $\mathcal{N}\subset[n]$, if $\textbf{a}^{\mathcal{N}} \in \textrm{Nash}\big(\Gamma_{\dagger} (\ba_\star^{-\mathcal{N}})\big) $ then $ (\textbf{a}^{\mathcal{N}} ,\ba_\star^{-\mathcal{N}} ) \in  \textrm{Nash}\big(\Gamma)$. That is, for any subset of players $\mathcal{N}$, if $\ba^\mathcal{N}$ is a Nash equilibrium of the induced subgame $\Gamma_{\dagger} (\ba_\star^{-\mathcal{N}})$, then $(\ba^{\mathcal{N}} ,\ba_\star^{-\mathcal{N}})$ constitutes a Nash equilibrium of $\Gamma$.
	
	\textbf{Condition 2:}
	$\exists\ba_\star\in\textrm{Nash}(\Gamma)$, $\forall \mathcal{N}\subset[n]$,  $\textrm{Nash}\big(\Gamma_{\dagger} (\ba_\star^{-\mathcal{N}})\big) = \{\ba_\star^{\mathcal{N}} \}$. That is, for any induced subgame $\Gamma_{\dagger} (\ba_\star^{-\mathcal{N}})$, the Nash equilibrium of $\Gamma_{\dagger} (\ba_\star^{-\mathcal{N}})$ is $\ba_\star^{\mathcal{N}}$, and is unique.
	
	Note that Condition 2 implies that $\ba_\star$ is strict, as it requires each 1-player subgame to have a unique equilibrium.
	
	The following is immediate:
	\begin{lemma}
		Condition 2 $\Rightarrow$ Condition 1 $\Rightarrow$ Condition 0. 
	\end{lemma}
	\begin{example} A trivial example of a game satisfying Condition 1 but not Condition 2 is a two-player game represented by the payoff matrix
		$$
		\begin{array}{cc}
			(1,1) & (1,0) \\ 
			(1,2) & (1,1)
		\end{array}
		$$
		
		This game has two pure Nash equilibrium which are not strict.\\
	\end{example}
	
	\begin{example}
		An example of a game satisfying Condition 0 but not Condition 1 is an $n$-player identical-interest game $\Gamma$ where for at least one $\mathcal{N}\subset[n]$, $\cap_{i\in\mathcal{N}} \textbf{A}^i\not=\emptyset$. 
		The common payoff for any $\textbf{a}=(a^1,\dots,a^n)\in\textbf{A}$ is given by the maximum number of players who choose the same action, so that the payoff function $r(\textbf{a})= \max_{b\in \textbf{B}}\sum_{i\in[n]} \mathbb{1}_{\{a^i=b\}}$ where $\textbf{B}:=\cup_{i\in[n]} A^i$. That is, the players' interests are fully aligned.
		
		Pick $\textbf{a}_1\in\textbf{A}$. Assume that $\textrm{Sat}(\textbf{a}_1)$ is a non-empty proper subset of $[n]$ (to avoid the trivial cases). 
		Let 
		\begin{eqnarray*}
			\mathcal{N}(b;\textbf{a}_1) & := & \{i\in[n]:a_1^i=b\}, \qquad \forall b\in\textbf{B} \\
			\textbf{B}_{\max} & := & \arg\max_{b\in\textbf{B}} |\mathcal{N}(b;\textbf{a}_1)| \\
			\bar{\textbf{B}}_{\max} 
			& := & \arg\max_{b\in\textbf{B}_{\max}}|\{i \in [n] \backslash \mathcal{N}(b;\textbf{a}_1) : b\in \textbf{A}^i\}|
		\end{eqnarray*}
		
		The set of pure Nash equilibria is
		$$\textrm{Nash}(\Gamma) = \cup_{b\in\textbf{B}}\{(b,\dots,b)\}.$$
	\end{example}
	
	\begin{theorem}\label{Cond0Thm}
		Condition 0 implies that $\Gamma$ is GenWAG.
	\end{theorem}
	
	\noindent\textbf{Proof.} Consider  $\textbf{a}_1\in\textbf{A}$. Assume that $\mathcal{N}_1:=\textrm{Sat}(\textbf{a}_1)$ is a non-empty proper subset of $[n]$ (to avoid the trivial cases). Let $\bar{\textbf{a}}\in\textrm{Nash}(\Gamma)$ correspond to $\textbf{a}_1$ in Condition~0. 
	Construct a satisficing a path starting at $\textbf{a}_1$ by the following steps.
	
	\begin{itemize}
		\item If $\mathcal{N}_1\not=[n]$, let
		$\textbf{a}_2:= (\textbf{a}_1^{\mathcal{N}_1},\bar{\textbf{a}}^{-\mathcal{N}_1})$ and  $\mathcal{N}_2:=\textrm{Sat}(\textbf{a}_2) \cap \mathcal{N}_1$.  
		\item If $\mathcal{N}_2\not=\mathcal{N}_1$, let
		$\textbf{a}_3:= (\textbf{a}_2^{\mathcal{N}_2},\bar{\textbf{a}}^{-\mathcal{N}_2})$ and  $\mathcal{N}_3:=\textrm{Sat}(\textbf{a}_3) \cap \mathcal{N}_2$.  
		\item $\dots$
		\item If $\mathcal{N}_k=\mathcal{N}_{k-1}$, stop.
	\end{itemize}
	This process ends  with a (finite) satisficing path $$\textbf{a}_1 \rightarrow\cdots\rightarrow \textbf{a}_k=(\textbf{a}_k^{\mathcal{N}_k},\bar{\textbf{a}}^{-\mathcal{N}_k})$$
	where
	$$ \mathcal{N}_{k-1}=\mathcal{N}_k \subset\textrm{Sat}(\textbf{a}_k)$$
	therefore
	$$\textbf{a}_k^{\mathcal{N}_k} \in \textrm{Nash}(\Gamma_{\dagger}(\bar{\textbf{a}}^{-\mathcal{N}_k})).$$
	Since $\mathcal{N}_k\subset\mathcal{N}_1=\textrm{Sat}(\textbf{a}_1)$,  we have, from Condition~0,
	$$\textbf{a}_k\in\textrm{Nash}(\Gamma).$$
	
	\hfill $\diamond $

	Observe that Condition 1 and 2 implies something stronger than being GenWAG: They require reachability of a specific equilibrium from every initial profile, whereas Condition 0 allows for initial profile dependence. Different applications may dictate different approaches. Interchangeable games appear to be a special case of Condition 0; As long as a subset of players is in equilibria, it doesn't matter which equilibrium the others are at.
	
	A useful special case of {\bf Condition 1} is given in the following:  
	
	\begin{corollary}\label{theorem:n-player-weaker-condition}
		Let $\Gamma$ be an $n$-player game, and suppose $\Gamma$ admits a strict pure Nash equilibrium $\ba_{\star}$. Further suppose that for each nonempty proper subset $\NN \subset[n]$ that the subgame $\Gamma_{\dagger}(\ba^{-\NN}_{\star})$ admits a unique pure Nash equilibrium. Then, the game $\Gamma$ satisfies Condition 1 and is therefore generalized weakly acyclic. 
	\end{corollary}

	The assumption in Corollary~\ref{theorem:n-player-weaker-condition} makes explicit use of satisficing dynamics by ensuring that if any player is at their component of $\ba_{\star}$, at least one other player who is \emph{not} at their component of $\ba_{\star}$ must be unsatisfied. Each unsatisfied player can then switch in turn to their component of $\ba_{\star}$, eventually leading to equilibrium. Additionally, uniqueness is not required in the full game $\Gamma$ (unlike in Condition 2), which allows for games with multiple pure Nash equilibria to be generalized weakly acyclic as long as one of them is strict and satisfies the condition given in Corollary \ref{theorem:n-player-weaker-condition}. To show that such a game is generalized weakly acyclic, one need only verify that for some strict $\ba_{\star}$ that each induced subgame $\Gamma^{\NN}_{\star}$ admits a unique pure Nash equilibrium. This is a much easier condition to verify than showing that this is true for \emph{all} induced subgames, as in Proposition \ref{theorem:n-player-sufficient-condition}. It is also easier to verify than Condition 0, which permits dependence on the initial action profile, or Condition 1, which allows for multiple subgame equilibria. One should also note that for $n=2$, Corollary~\ref{theorem:n-player-weaker-condition} only requires that each single-player subgame possess a strict Nash equilibrium, as uniqueness is not required in the full game. This retrieves the sufficiency condition given in Lemma \ref{theorem:two-player-sufficient-condition}, as expected.

	\section{Stochastic Dynamic GenWAGs: Paths on Normal Form Strategies}
	
	By viewing stationary policies as {\it actions} in a lifted game in normal form, in this section we discuss the generalization of the results presented above in the context of stochastic dynamic games. For clarity, in the following sections we use the terms "DM" and "agent" somewhat interchangeably to refer to entities that select actions according to policies. We also adopt the standard stochastic control convention of minimizing a cost function, rather than maximizing a reward function. These will be utilized in the section to follow where learning results will be presented.
	
	Consider a finite discounted stochastic game ${\cal G}$:
	\begin{description}
		\item[(i)] A finite set $[n]=\{1,\dots,n\}$ of DMs
		\item[(ii)] a finite set $\mathbb{X}$ of states
		\item[(iii)] a finite set $\mathbb{U}^i$ of control decisions for each DM$i$
		\item[(iv)] a cost function $c^i$ for each DM$i$, representing DM$i$'s cost $c^i(x,u^1,\dots,u^n)$ at each state $x\in\mathbb{X}$ and for each joint decision $(u^1,\dots,u^n)\
		\in\mathbb{U}^1\times\cdots\mathbb{U}^n$
		\item[(v)] a discount factor $\beta^i\in(0,1)$ for each DM$i$
		\item[(vi)] a random initial state $x_0\in\mathbb{X}$
		\item[(vii)] a transition kernel 
		$P[ x^{\prime} |x,u^1,\dots,u^n]$ for the probability of each state transition from $x\in\mathbb{X}$ to $x^{\prime}\in\mathbb{X}$ for each joint decision $(u^1,\dots,u^n)\in\mathbb{U}^1\times\cdots\mathbb{U}^n.$
	\end{description}
	
	Such a stochastic game induces a discrete-time controlled
	Markov process where the state at time $t$ is
	denoted by $x_t\in\mathbb{X}$ starting with the initial state $x_0$.
	At any time $t\geq0$, each DM$i$ makes a control decision
	$u_t^i\in \mathbb{U}^i$ (possibly randomly) based on the available information. The
	state $x_t$ and the joint decisions $(u_t^1,...,u_t^n)$
	together determine each DM$i$'s cost $c^i(x_t,u_t^1,...,u_t^n)$ at time $t$ as well as the probability
	distribution $P[ \ \cdot \ | \ x_t,u_t^1,...,u_t^n]$ with which the next state $x_{t+1}$ is
	selected. We will focus on stationary policies where
	a DM's decision at time $t$ is determined solely based on the state $x_t$. We will denote the set of such policies by $\Gamma^i_S$ for each DM$i$. In this section, we will primarily be interested in deterministic (stationary) policies, denoted by $\Pi^i$ for each DM$i$, where each  policy $\pi^i\in\Pi^i$ is identified by a mapping from $\mathbb{X}$ to $\mathbb{U}^i$. 
	
	Instead of attempting to maximize a reward function as in the prior sections, here the objective of each DM$i$ is to find a policy
	$\pi^i\in\Gamma^i_S$ that minimizes its expected discounted  cost
	\begin{equation}
		J^i_{x}(\pi^1,\dots\pi^n)   =  E_{x}[\sum_{t\geq0} (\beta^i)^t c^i\left(x_t,u^1_t,\dots,u^n_t\right)] \label{eq:dc}
	\end{equation}
	for all $x\in\mathbb{X}$, where $E_{x}$ denotes the conditional expectation given  $x_0=x$.
	Since DMs have possibly different cost functions and each DM's
	cost may depend on the control decisions of the other DMs, we adopt
	the notion of equilibrium  to
	represent those policies that are {\it person-by-person optimal} (or equivalently, at a {\it Nash equilibrium}). For ease of
	notation, we denote the policies of all DMs other than DM$i$ by $\pi^{-i}$. We also define ${\bf \Pi}^{-i}:=\times_{j\not=i} \Pi^j$ and ${\bf \Gamma}^{-i}_S:=\times_{j\not=i} \Gamma^j_S$ as well as ${\bf \Pi}:=\times_j \Pi^j$ and $\Delta:=\times_j \Gamma^j_S$. Using this notation, we write a joint policy
	$(\pi^1,\dots\pi^n)$ as $(\pi^{i},\pi^{-i})$ and $J_x^i(\pi^1,\dots\pi^n)$
	as $J_x^i(\pi^{i},\pi^{-i})$. A joint policy $(\pi^{1}_\star,\dots,\pi^{n}_\star)\in\Delta$ constitutes an (Markov perfect) equilibrium (see \citet[Chapter 10]{YukselBasarBook24}) if, for all $i$, $x$, $$J_x^i(\pi^{i}_\star,\pi^{-i}_\star)=\min_{\pi^i\in\Gamma^i_S}J_x^i(\pi^{i},\pi^{-i}_\star).$$
	It is known that any finite discounted stochastic game possesses an equilibrium policy as defined above. Although the minimum above can always be achieved by a deterministic policy in $\Pi^i$ (since each DM$i$'s problem is a stationary Markov decision problem when the policies of the other DMs are fixed at $\pi^{-i}_\star$), a deterministic equilibrium policy may not exist in general. However, many interesting classes of games do possess equilibrium in deterministic policies. 
	
	In particular, large classes of games arising from applications where all DMs benefit from cooperation possess equilibrium in deterministic policies \citet{YukselBasarBook24}. The primary examples of such games of cooperation are team problems where all DMs have the same cost function, i.e., $c^i=c$ for all $i$. In such team problems, deterministic policies minimizing the common cost function are clearly equilibrium policies although non-optimal deterministic equilibrium policies may also exist. A more general set of games of cooperation are those in which some function, called the potential function, decreases whenever a single DM decreases its own cost by unilaterally switching from one deterministic policy to another one. In this class of games, the deterministic policies minimizing the potential function are equilibrium policies.  It should be noted that both potential games and team games are in the class of weakly acyclic games; such games have recently been receiving much attention \citet{guo2025markov,guo2025towards,arefizadeh2024characterizations,leonardos2021global,hosseinirad2026linear} due to their mathematical convenience in studying stochastic games. As such, in this context, we are primarily interested in the set of deterministic equilibrium policies denoted by ${\bf \Pi}_{\rm eq}$, where ${\bf \Pi}_{\rm eq}\subset {\bf \Pi}$. 
	
	To support the application of our results to stochastic dynamic games, we review the setup used in \citet{ArslanYukselTAC16} for weakly acyclic games. As studied earlier, with deterministic stationary policies serving as the actions, we have the following discussion.
	\subsection{Weakly Acyclic Games}
	Let $\Pi^i_{\pi^{-i}}$ denote DM$i$'s set of (deterministic) best replies to any $\pi^{-i}\in{\bf \Gamma}^{-i}_S$, i.e.,
	\begin{align*}
		\Pi_{\pi^{-i}}^i  := \big\{\hat{\pi}^i\in\Pi^i  :  \ & J_x(\hat{\pi}^i,\pi^{-i})=\min_{\pi^i\in\Gamma^i_S}J_x(\pi^i,\pi^{-i}), \quad \mbox{for all} \  x \big\}.
	\end{align*}
	DM$i$'s best replies to any $\pi^{-i}\in{\bf \Gamma}^{-i}$ can be characterized by its optimal Q-factors $Q_{\pi^{-i}}^i$ satisfying the fixed-point equation
	\begin{align}
		\nonumber
		Q_{\pi^{-i}}^i(x,u^i) = & E_{\pi^{-i}(x)} \big[c^i(x,u^i,u^{-i}) +\beta^i \sum_{x^{\prime}\in\mathbb{X}} P[x^{\prime}|x,u^i,u^{-i}]\min_{v^i\in\mathbb{U}^i} Q_{\pi^{-i}}^i(x^{\prime},v^i) \big]
		\label{eq:Qfp}
	\end{align}
	for all $x,u^i$, where $E_{\pi^{-i}(x)}$ denotes the expectation with respect to the joint distribution of $u^{-i}$ given by $\pi^{-i}(x)=\pi^1(x)\times\cdots\times\pi^{i-1}(x)\times\pi^{i+1}(x)\times\cdots\times\pi^n(x)$. The optimal Q-factor $Q_{\pi^{-i}}^i(x,u^i)$ represents DM$i$'s expected discounted cost-to-go from the initial state $x$ assuming that DM$i$ initially chooses $u^i$ and uses an optimal policy thereafter while the other DMs use $\pi^{-i}$.
	One can then write  $\Pi_{\pi^{-i}}^i$ as
	\begin{align*}
		\Pi_{\pi^{-i}}^i = \big\{\hat{\pi}^i\in\Pi^i: Q_{\pi^{-i}}^i(x,\hat{\pi}^i(x))=\min_{v^i\in\mathbb{U}^i} Q_{\pi^{-i}}^i(x,v^i), \quad \mbox{for all} \ x \big\}.
	\end{align*}
	The set of (deterministic) joint best replies is denoted by ${\bf \Pi}_{\pi}:=\Pi_{\pi^{-1}}^1\times\cdots\times\Pi_{\pi^{-n}}^n$.
	Any best reply $\hat{\pi}^i\in\Pi_{\pi^{-i}}^i$ of DM$i$ is called a {\it strict best reply} with respect to $(\pi^i,\pi^{-i})$ if
	\begin{align*}
		J_x^i(\hat{\pi}^i,\pi^{-i}) & < J_x^i(\pi^i,\pi^{-i}), \quad\mbox{for some} \ x.
	\end{align*}
	Such a strict best reply $\hat{\pi}^i$ achieves DM$i$'s minimum cost  given $\pi^{-i}$ for all initial states, and results in a strict improvement over $\pi^i$ for at least one initial state.
	\begin{definition}\label{DefinitionSBRPP}
		We call a (possibly finite) sequence of deterministic joint policies $\pi_0,\pi_1,\dots$  a {\it strict best reply path} if, for each $k$, $\pi_k$ and $\pi_{k+1}$ differ in exactly one DM position, say DM$i$, and $\pi_{k+1}^i$ is a strict best reply with respect to $\pi_{k}$.
	\end{definition}
	\begin{definition}\label{def:best}
		A discounted stochastic game is called \\ {\it weakly acyclic} under strict best replies if there is a strict best reply path starting from each deterministic joint policy and ending at a deterministic equilibrium policy.
	\end{definition}
	
	
	%
	%
	
	Figure~\ref{fig33} shows the policy-revision graph of a game where the nodes represent the deterministic joint policies and the directed edges represent the revision paths. According to Definition~\ref{def:best}, a game is weakly acyclic if, from any starting joint policy, there exists a finite best-response path to equilibrium. In this graph, this is visualized as paths following the directed edges, with equilibrium policies represented by sinks in the graph.

	\subsection{Stochastic Dynamic Games with Pure Stationary Policies which are GenWAG but not Weakly Acyclic.}
	\begin{definition}\label{DefinitionSSPP}
	We call a (possibly finite) sequence of deterministic joint policies $\pi_0,\pi_1,\dots$  a {\it satisficing path} if, for each $k$, $\pi_k$ and $\pi_{k+1}$ differ in exactly one DM position, say DM$i$, where $\pi_{k+1}^i$ is any policy.
	\end{definition}
	\begin{definition}\label{def:satisficing}
	A discounted stochastic game is called {\it generalized weakly acyclic} if there is a satisficing path starting from each deterministic joint policy and ending at a deterministic equilibrium policy.
	\end{definition}
	
	Definition~\ref{DefinitionSSPP} and Definition~\ref{def:satisficing} provide clear extensions of Definition~\ref{DefinitionSBRPP} and Definition~\ref{def:best} to GenWAGs, where unsatisfied DMs can now update to \emph{any} policy, not just a strict best reply. Having shown earlier that the set of GenWAGs is a consequential structural generalization of weakly acyclic games, we provide an explicit example in the context of stochastic Markov games, where policies are stationary. We note 3 classes of game that are GenWAG:
	\begin{itemize}
	\item All symmetric games with a pure Nash equilibrium; see Theorem \ref{thm:symmetric-paths} \citet{yongacoglu2021satisficing} which directly applies to such a setting as well.
	\item Weakly Acyclic Games, including potential games and team games
	\item Games that satisfy \textbf{Condition 0}
	\end{itemize}
	
	Furthermore, Theorem \ref{Cond0Thm} applies identically to the normal form dynamic game setting as well. 
	
	
	\begin{example}\label{example:stochasticGenwag}
	The following is a pure-strategy stochastic game that is generalized weakly acyclic but not weakly acyclic.
	\begin{figure}[h]
		\centering
		\begin{subfigure}[b]{0.3\textwidth}
			\centering
			{\small
				\begin{game}{3}{3}[$s_1$]
					& $L$       & $C$       & $R$ \\
					$T$     & 0, 0      & 9, 9       & 9, 9 \\
					$M$     & 9, 9      & 8, 9       & 9, 8 \\
					$B$     & 9, 9      & 9, 8       & 8, 9 \\
				\end{game}
			}
		\end{subfigure}
		\begin{subfigure}[b]{0.3\textwidth}
			\centering
			{\small
				\begin{game}{2}{2}[$s_2$]
					& $W$       & $Z$       \\
					$X$     & 9, 9      & 8, 8     \\
					$Y$     & 8, 8      & 9, 9      \\
				\end{game}
			}
		\end{subfigure}
		\caption{A stochastic game that is not weakly acyclic but is generalized weakly acyclic} \label{fig:non-acyclic-markov-game}
		\label{fig:stochastic-GenWAG}
	\end{figure}
	
	\begin{figure}[h]
		\begin{subfigure}[h]{0.75\textwidth}
			{\small
				\begin{game}{6}{6}[$x_0=s_1$]
					&$(L,W)$&$(L,Z)$&$(C,W)$&$(C,Z)$&$(R,W)$&$(R,Z)$\\
					$(T,X)$   &(0,0)&(0,0)&(18,18)&(18,18)&(18,18)&(18,18)\\ 
					$(T,Y)$   &(0,0)&(0,0)&(18,18)&(18,18)&(18,18)&(18,18)\\
					$(M,X)$   &(18,18)&(18,18)&(16 2/3,18)&(16,17)&(18,14 2/3)&(17,16)\\
					$(M,Y)$   &(18,18)&(18,18)&(16,17)&(17,18)&(17,16)&(18,17)\\
					$(B,X)$   &(18,18)&(18,18)&(18,14 2/3)&(17,16)&(16 2/3,18)&(16,17)\\
					$(B,Y)$   &(18,18)&(18,18)&(17,16)&(18,17)&(16,17)&(17,18)\\
				\end{game}
			}
		\end{subfigure}
		\begin{subfigure}[h]{0.75\textwidth}
			{ \small
				\begin{game}{6}{6}[$x_0=s_2$]
					&$(L,W)$&$(L,Z)$&$(C,W)$&$(C,Z)$&$(R,W)$&$(R,Z)$\\
					$(T,X)$   &(9,9)&(16,16)&(18,18)&(16,16)&(18,18)&(16,16)\\ 
					$(T,Y)$   &(16,16)&(18,18)&(16,16)&(18,18)&(16,16)&(18,18)\\
					$(M,X)$   &(18,18)&(16,16)&(17 1/3,18)&(16,16)&(18,17 1/3)&(16,16)\\
					$(M,Y)$   &(16,16)&(18,18)&(16,16)&(18,18)&(16,16)&(18,18)\\
					$(B,X)$   &(18,18)&(16,16)&(18,17 1/3)&(16,16)&(17 1/3,18)&(16,16)\\
					$(B,Y)$   &(16,16)&(18,18)&(16,16)&(18,18)&(16,16)&(18,18)\\
				\end{game}
			}
		\end{subfigure}
		\caption{Expected discounted reward starting from $s_1$ and $s_2$}
		\label{fig:stochastic-GenWAG-expected-reward}
	\end{figure}
	
	Each player has a discount factor of $\beta=0.5$, and transition probabilities are as follows: $P[s_2|s_1,u^1,u^2]=1$ only when $(u^1,u^2)\in\{M,B\}\times\{C,R\}$, with $P[s_2|s_1,u^1,u^2]=0$ otherwise. $P[s_1|s_2,u^1,u^2]=1$  when $(u^1,u^2)=(X,W)$, with $P[s_1|s_2,u^1,u^2]=0$ for any other $(u^1,u^2)$.
	
	To see why this game is not weakly acyclic but is GenWAG, we look at the expected discounted cost for each joint policy. Each row in Figure \ref{fig:stochastic-GenWAG-expected-reward} corresponds to the policy $\pi^1 =(u_{s_1}^1,u_{s_2}^1)$ for player 1, and each column corresponds to $\pi^2 =(u_{s_1}^2,u_{s_2}^2)$ for player 2. The expected discounted cost for player 1 and 2 is then written in each entry as $(J^1_{x_0},J^2_{x_0})$.
	
	Consider the starting strategy $\pi=((M,X),(C,Z))$. Player 1 is satisfied for either choice of $x_0$, but player 2 is not for $x_0=s_1$. As they are not allowed to increase their cost in $s_2$, their only choice is to choose $(R,Z)$. From here player~2 is now satisfied, and player 1 is unsatisfied, and can only choose to play $(B,X)$. Player 2 is then left with no choice but to play $(C,Z)$, and player 1 is compelled by better-response dynamics to play $(M,X)$, which brings us back to the initial strategy. In other words, this is an inescapable cycle in the better-response graph. On the other hand, it is easy to see that satisficing dynamics allow any initial strategy to reach the equilibrium $((T,X),(L,W))$.
	\end{example}

	\section{Implications for Multi-Agent Reinforcement Learning (MARL) and Stochastic Markov Games}
	
	In this section, we discuss and reflect on the implications on our analysis above on learning in stochastic dynamic games. Notably, the analysis shows that via satisficing paths, and by viewing the stationary policy maps as the normal form strategies used in the analysis in the previous sections, one can significantly enlarge the class of games for which a two-time scale paradigm can be applied towards convergence to equilibria. 
	
	Reinforcement learning algorithms use past data to design control policies; we refer the reader to studies on Q-learning \citet{Watkins}, \citet{TsitsiklisQLearning}, \citet{bormey00a} and several comprehensive references \citet{CsabaAlgorithms} and \citet{meyn2022control}. For partially observable models, a detailed review is available in  \citet{singh1995reinforcement,kara2021convergence}.
	
	
	The study of decentralized systems is known to be challenging both for stochastic teams and stochastic games. There are two primary challenges in learning for such systems: (i) the first immediate challenge for learning in such models is due to decentralization of information, which prevents the standard approaches used in fully or partially observable MDPs to be inapplicable; (ii) the second difficulty comes from the non-stationarity of the environment from the point of view of any individual agent. As an agent learns how to improve its performance, it will alter its behavior, and this can have a destabilizing effect on the learning processes of the remaining agents, who may change their policies in response to outdated strategies. Notably, this issue arises when one tries to apply single-agent reinforcement-learning algorithms in multi-agent settings. A number of studies have reported on non-convergent play when single-agent algorithms using local information are employed, without modification, in multi-agent settings. Thus, for such models a primary obstacle to convergence of Q-learning is due to the presence of multiple active learners leading to a non-stationary environment for all learners.  
	
	\subsection{A General Paradigm for Multi-Agent Learning: Two-Time Scales and a Markov Chain over Play Path Graph}\label{se:lp}
	
	To overcome the obstacles noted above, building on prior work \citet{foster2006regret, germano2007global}, \citet{ArslanYukselTAC16} modified the Q-learning for stochastic games as follows: In the variation of Q-learning, DMs are allowed to use constant policies for extended periods of time called {\it exploration phases}. This is sometimes referred to as a {\it two-time scales} approach\footnote{We note that an alternative multiple time scales approach is via the different learning rates applied by agents by taking advantage of slow learning (typically of policies) and fast learning (typically of values); see \citet{borkar1997stochastic,borkar2002reinforcement,leslie2003convergent,leslie2005individual,sayin2021decentralized}. This approach couples the stochastic learning dynamics with a system of ODEs which characterizes best-response dynamics, whose stability can be used to establish convergence in a variety of setups.}, and is the approach we consider here.
	
	The learning setup involves specifying the information that DMs have access to. To make the discussion more concrete, let us first consider the following model. Suppose that each DM$i$ knows its own set of decisions $\mathbb{U}^i$ and its own discount factor $\beta^i$. In addition, before choosing its decision $u_t^i$ at any time $t$, each DM$i$ may have more specific information. In particular, if the DMs have access to the global state information, we specify the following information structure for DM$i$, for time $t$ (this will be relaxed later): (i) its own past decisions $u_0^i,\dots,u_{t-1}^i$, (ii) past and current state realizations $x_0,\dots,x_t$, and (iii) its own past cost realizations $$c^i(x_0,u_0^i,u_0^{-i}),\dots,c^i(x_{t-1},u_{t-1}^i,u_{t-1}^{-i}).$$
	
	Each DM$i$ has access to no other information such as the state transition probabilities or any information regarding the other DMs (not even the existence of such DMs). In effect, the problem of decision making from the perspective of each DM$i$ appears to be a stationary Markov decision problem. It is reasonable that each DM$i$ with this view of its environment would use the standard Q-learning algorithm \citet{TsitsiklisQLearning} to learn its optimal Q-factors and its optimal decisions. This would lead to the following Q-learning dynamics for each DM$i$:
	\begin{align*}
		Q_{t+1}^i(x,u^i)  = & Q_t^i(x,u^i), \quad \mbox{for all} \ (x,u^i)\not= (x_t,u_t^i)\\
		Q_{t+1}^i(x_t,u_t^i)  = & Q_t^i(x_t,u_t^i) + \alpha_t^i (c^i(x_t,u_t^i,u_t^{-i}) \\&+\beta^i \min_{v^i\in\mathbb{U}^i} Q_t^i(x_{t+1},v^i)-Q_t^i(x_t,u_t^i))
	\end{align*}
	where $\alpha_t^i\in[0,1]$ denotes DM$i$'s step size at time $t$, and $u_t^{-i}$ denotes the actions of all DMs except DM$i$.
	
	If only one DM, say DM$i$, were to use Q-learning and the other DMs used constant policies $\pi^{-i}$, then DM$i$ would asymptotically learn its corresponding optimal Q-factors, i.e., $P(Q_t^i \rightarrow Q_{\pi^{-i}}^i)=1$ provided that all state-control pairs $x,u^i$ are visited infinitely often and the step sizes are reduced at a proper rate. This follows from the convergence of Q-learning in a stationary environment. 
	
\begin{figure}[H]
		\centering
	\epsfig{figure=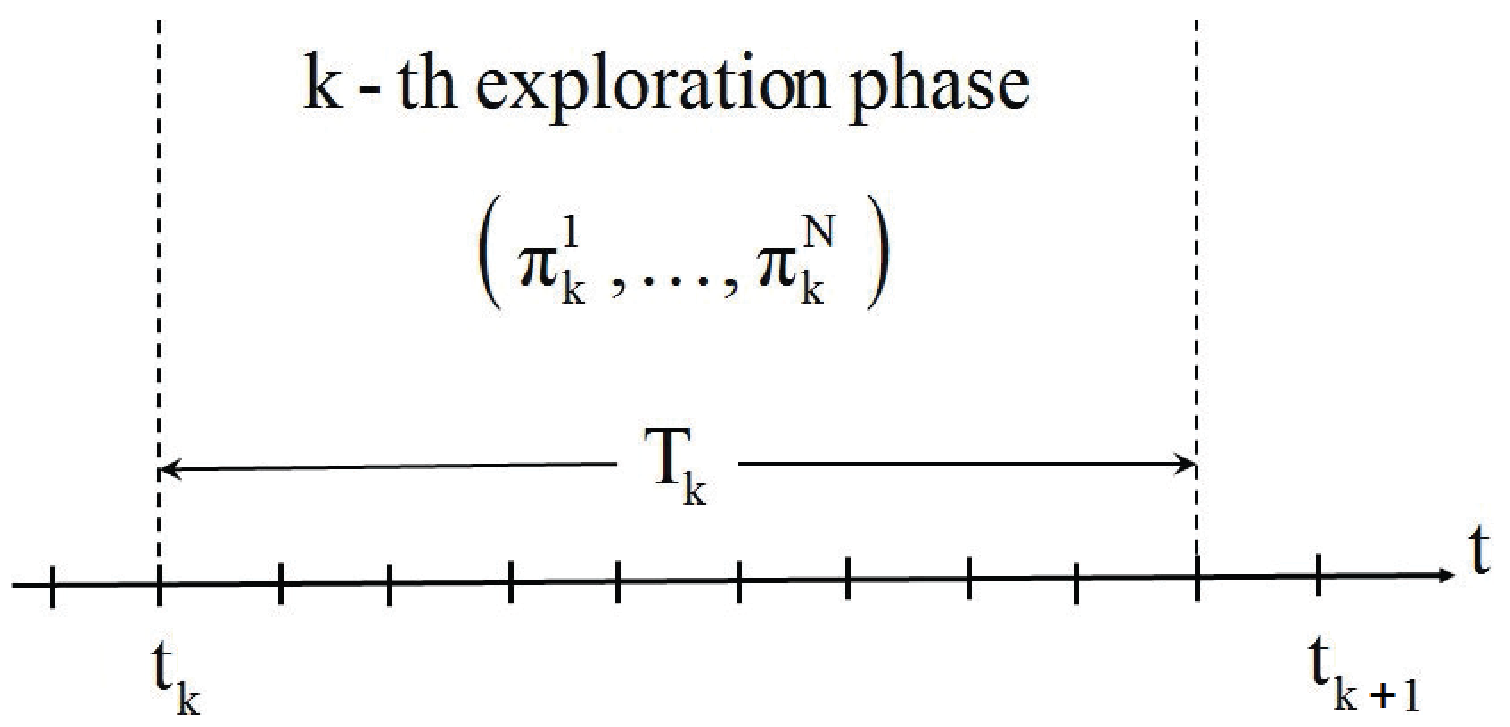,height=2.1cm,width=5cm}
	\caption[]{An illustration of the $k-$th exploration phase.} \label{fig5}
	\end{figure}

	However, when all DMs use Q-learning and select their decisions as described above, the environment is non-stationary for all DMs, and there is no reason to expect convergence in that case. To overcome this obstacle, in the variant of learning presented in \citet{ArslanYukselTAC16}, the DMs are allowed to use constant policies for extended periods of time called {\it exploration phases}. As illustrated in Figure~\ref{fig5}, the $k-$th exploration phase runs through times $t=t_k,\dots,t_{k+1}-1$, where \[t_{k+1}=t_k+T_k \qquad \mbox{(with $t_0=0$)}\] for some integer $T_k\in[1,\infty)$ denoting the length of the $k-$th exploration phase. During the $k-$th exploration phase, DMs use some constant policies $\pi_k^1,\dots,\pi_k^N$ as their baseline policies with occasional experimentation.   
	
\begin{figure}[H]
	\centering
	\epsfig{figure=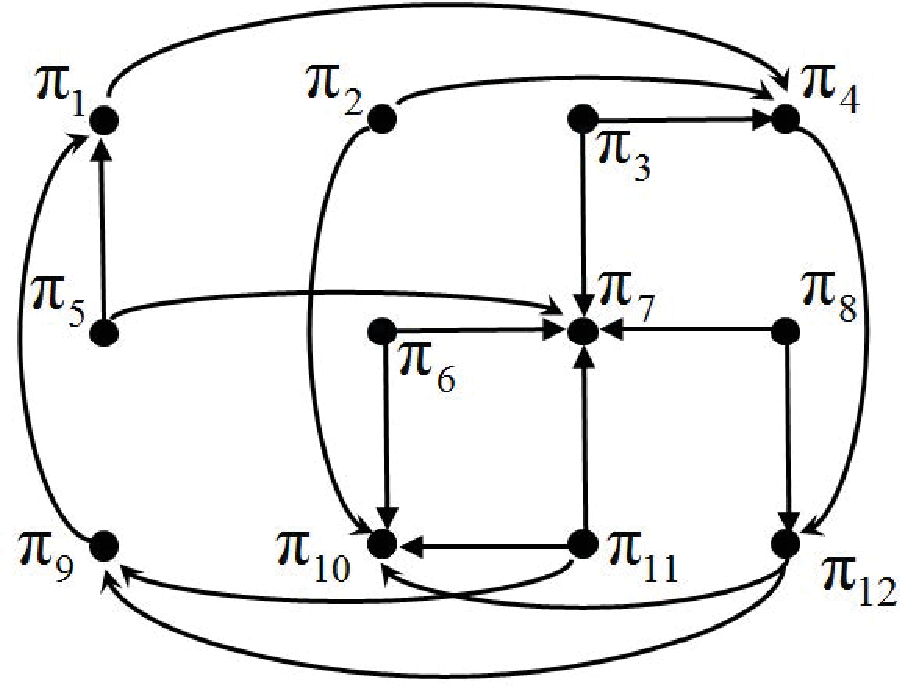,height=3.2cm,width=4.0cm}
	\caption[]{The response graph of a stochastic game.} \label{fig33}
\end{figure}

	The main idea is to create a stationary environment over each exploration phase so that DMs can almost accurately learn their optimal Q-factors corresponding to the constant policies used during each exploration phase and update their policies; see Figure \ref{fig33} for an example of a response graph of a stochastic game, where $\pi_7$ is an absorbing strategy.
	
\subsection{Algorithms and Convergence.} 

Let us now describe a generalization of the algorithm introduced in \citet{ArslanYukselTAC16} that allows for learning in GenWAGs, a setting not covered by the original algorithm.

\begin{algorithm}[!b]
\caption{Learning for GenWAGs with Local Action Information}
\label{al:2GenWAGPure} 
[for DM$^i$]

\noindent Set parameters\\
\hspace*{5mm} $\mathbb{Q}^i$: some  compact subset in $\mathbb{R}^{|\mathbb{X}\times\mathbb{U}^i|}$ \\
\hspace*{11mm} where $|\mathbb{X}\times\mathbb{U}^i|$ is the number of pairs $(x,u^i)$\\
\hspace*{5mm} $\{T_k\}_{k\geq0}$: sequence of integers in $[1,\infty)$ \\
\hspace*{5mm} $\rho^i\in(0,1)$: experimentation probability\\
\hspace*{5mm} $e^i\in(0,1)$: random policy updating probability\\
\hspace*{5mm} $\delta^i\in(0,\infty)$: tolerance level for sub-optimality\\
\hspace*{5mm} $\{\alpha_{v}^{i}\}_{v\geq0}$:  sequence  of step sizes where \\
\hspace*{15mm} $\alpha_{v}^{i}\in[0,1]$, $\sum_v\alpha_v^{i} = \infty$, $\sum_{v}\big(\alpha_v^{i}\big)^2 < \infty$ \\
\hspace*{15mm}    (e.g., $\alpha_v^i=1/v^r$ where $r\in(1/2,1]$)

\noindent Initialize  $\pi_0^i \in \Pi^i$ (arbitrary), $Q_0^i\in\mathbb{Q}^i$ (arbitrary) \\
Receive $x_0$ \\
Iterate $k\geq0$ \\
\hspace*{5mm}($k-$th exploration phase)\\
\hspace*{5mm}Iterate $t =t_k,\dots,t_{k+1}-1$ \\
\hspace*{10mm}  $u_t^i = \left\{\begin{array}{cl} \pi_k^i(x_t), & \textrm{ w.p. } 1-\rho^i\\ \textrm{any } u^i\in\mathbb{U}^i, & \textrm{ w.p. } \rho^i/|\mathbb{U}^i| \end{array} \right.$ \\
\hspace*{10mm}    Receive $c^i(x_t,u_t^i,u_t^{-i})$ \\
\hspace*{10mm}    Receive $x_{t+1}$ (selected according to  $P[ \ \cdot \ | \ x_t,u_t^i,u_t^{-i}]$) \\
\hspace*{10mm}    $v_t^i =$ the number of visits to $(x_t,u_t^i)$ in the \\
\hspace*{10mm}     $k-$th exploration phase up to $t$ \\
\hspace*{10mm} $Q_{t+1}^i(x_t,u_t^i)  =   (1-\alpha_{v_t^i}^{i})Q_t^i(x_t,u_t^i)$ \\
\hspace*{20mm} $ + \alpha_{v_t^i}^{i} \big[ c^i(x_t,u_t^i,u_t^{-i})  + \   \beta^i \min_{v^i} Q_t^i(x_{t+1},v^i) \big]$\\
\hspace*{10mm}    $Q_{t+1}^i(x,u^i) =   Q_t^i(x,u^i)$,   for all $(x,u^i)\not=(x_t,u_t^i)$\\
\hspace*{5mm} End \\
\hspace*{5mm} $\Pi_{k+1}^i  = \big\{\hat{\pi}^i\in \Pi^i:  Q_{t_{k+1}}^i(x,\hat{\pi}^i(x))$ \\
\hspace*{20mm} $\leq \min_{v^i}Q_{t_{k+1}}^i(x,v^i)+\delta^i, \ \mbox{for all}  \ x\big\}$ \\
\hspace*{5mm} If  $\pi_k^i \in \Pi_{k+1}^i$ \\
\hspace*{10mm} $\pi_{k+1}^i = \pi_k^i$ \\
\hspace*{5mm} Else \\
\hspace*{10mm} $\pi_{k+1}^i = \left\{\begin{array}{cl} \pi^i_k& \textrm{ w.p. } 1-e^i \\\textrm{any }\pi^i\in\Pi^i, & \textrm{ w.p. } e^i/|\Pi^i|  \end{array} \right.$
\\
\hspace*{5mm} End \\
\hspace*{5mm} Reset $Q_{t_{k+1}}^i$ to any $Q^i\in\mathbb{Q}^i$ (e.g., project $Q_{t_{k+1}}^i$ onto $\mathbb{Q}^i$)\\
End
\end{algorithm}

Each DM using Algorithm~\ref{al:2GenWAGPure} approximately learns its optimal Q-factors during each exploration phase with limited observations. Accordingly, each DM updates its (baseline) policy to one of its near best replies with inertia based on its learnt Q-factors. Hence, Algorithm~\ref{al:2GenWAGPure} can be regarded as an approximation to a best reply process (with inertia).


We require the following assumption.

\begin{assumption}
\label{as:alpha}
For all $(x^{\prime},x)$, there exists a finite integer $H\geq0$ and joint actions $\tilde{u}_0,\dots,\tilde{u}_{H}$ such that \[P \bigg(x_{H+1}=x^{\prime} \ | \ (x_0,u_0,\dots,u_{H})=(x,\tilde{u}_0,\dots,\tilde{u}_{H})\bigg)>0.\]
\end{assumption}

This assumption ensures that for any state $x$, it is possible to reach any state $x^\prime$ in a finite number of steps $H$ via some sequence of joint actions $\tilde{u}_0,\dots,\tilde{u}_{H}$. Without this assumption, there could exist some initial state $x_0=x$ from which $x'$ is unreachable in finite time, which would impede the ability to learn over all states, and thus convergence.

\begin{theorem}\label{normalQgenWAG}
Consider a discounted stochastic game that is generalized weakly acyclic. Suppose that  each DM$i$ updates its policies by Algorithm~\ref{al:2GenWAGPure}. Let Assumption \ref{as:alpha} hold.
\begin{description}
\item[(i)] For any $\epsilon>0$, there exist $\tilde{T}<\infty$, $\tilde{k}<\infty$ such that if $\min_{\ell}T_{\ell}\geq \tilde{T}$,  then
$$P(\pi_{k} \in {\bf \Pi}_{\rm eq}) \geq 1-\epsilon, \qquad \mbox{for all} \ k\geq\tilde{k}.$$
\item[(ii)]  If $T_k \rightarrow \infty$, then $$P(\pi_{k} \in {\bf \Pi}_{\rm eq}) \rightarrow 1.$$
\item[(iii)] There exist finite integers $\{\tilde{T}_k\}_{k\geq0}$ such that if $T_k\geq\tilde{T}_k$, for all $k$, then
$$ P(\pi_{k} \rightarrow \pi_\star, \ \mbox{for some} \ \pi_\star\in{\bf \Pi}_{\rm eq}) = 1.$$
\end{description}
\end{theorem}

We now discuss the main insight behind this result. As all DMs use constant policies throughout any particular exploration phase, each DM indeed faces a stationary MDP in each exploration phase. Therefore, if the length of each exploration phase is long enough and the experimentation probabilities $\rho^1,\dots,\rho^N$ are small enough (but non-zero), each DM$i$ can learn its corresponding optimal Q-factors in each exploration phase with arbitrary accuracy with arbitrarily high probability. This allows each DM$i$ to accurately compute its near best replies to the other DMs' policies $\pi_k^{-i}$ at the end of the $k$-th exploration phase. By allowing each DM$i$ to update its policy $\pi_k^i$ with some probability $e^i\in(0,1)$ when not playing a near best reply (to $\pi_k^{-i}$), the resulting policy adjustment process intuitively approximates a satisficing reply process with inertia.

Note that since the players are learning over pure strategies, small perturbations will not alter the paths: the discrete structure is robust to small changes. The following argument mirrors that used in \citet{ArslanYukselTAC16}, which proves convergence of their algorithm for weakly acyclic games. The only difference is in the proof of {\it Part (i)}, where generalized weak acyclicity guarantees the existence of a satisficing path to an equilibrium policy from each starting policy, rather than a strict best reply path. We provide a sketch for completeness.

\noindent \textbf{Proof.} During the $k-$th exploration phase, each DM$^i$ actually uses the random policy $\bar{\pi}_k^{i}$ defined as
\begin{align}
\label{eq:pibar}
\bar{\pi}_k^j=(1-\rho^j)\pi_k^j+\rho^j\nu^j
\end{align}
where $\nu^j$ is the random policy that assigns the uniform distribution on $\mathbb{U}^j$ to each $x$. Let $\bar\delta$ denote the minimum separation between the entries of DMs' optimal Q-factors (for deterministic policies). We consider $\bar\delta$ to be the upper bound on the tolerance levels for sub-optimality, so that $\delta^i\in(0,\bar\delta)$ for all $i$. By \citet[Lemma 3]{ArslanYukselTAC16} there exists an upper bound $\bar\rho>0$ on the experimentation rates such that if $\rho^i\leq\bar\rho$  for all $i$, then
\begin{equation}
\label{eq:condrho}
\left|Q_{\pi_k^{-i}}^i - Q_{\bar{\pi}_k^{-i}}^{i}\right|_{\infty} < \frac{1}{2}\min\{\delta^i,\bar{\delta}-\delta^i\}, \quad\mbox{for all} \ i, \ k.
\end{equation}
In other words, we can always find $\bar\rho$ small enough to ensure that the optimal Q-factors for  $\bar\pi^{-i}_k$ remain close enough to those for $\pi^{-i}_k$, for all $i$. \citet[Lemma 2]{ArslanYukselTAC16} and (\ref{eq:condrho}) provide for the existence of some $\bar{T}<\infty$ such that if $T_k\geq\bar{T}$ then $P\left[ E_k  \right] \geq 1-\epsilon$ where $E_k$, $k\geq0$, is the event defined as
\begin{align*}
E_k:=\Big\{\omega\in\Omega:  \left|Q_{t_{k+1}}^i - Q_{\pi_k}^{i}\right|_{\infty} < & \frac{1}{2}\min\{\delta^i,\bar{\delta}-\delta^i\}, \ \forall i   \Big\}.
\end{align*}
That is, one can ensure at the end of exploration period $k$ that the learned Q-factor $Q_{t_{k+1}}^i$ has arbitrarily high probability of being within $\frac{1}{2}\min\{\delta^i,\bar{\delta}-\delta^i\}$ of the optimal Q-factor for every $(x,u^i)$.

{\it Part (i) \citet[Appendix B]{ArslanYukselTAC16}} Without loss of generality, assume $\epsilon\in(0,1)$. Because we have a generalized weakly acyclic game, there exists a satisficing path of minimum length $L_\pi<\infty$ starting at $\pi$ and ending at equilibrium for each $\pi\in\Pi$. Let $L:=\max_{\pi\in\Pi}L_\pi$. There then exists some probability $p_{\min}$ dependent only on the random policy update probability $\{e^i\}_{i=1}^n$ and $L$ such that for all $k$,
\begin{equation}
P\big[  \pi_{k+L} \in \Pi_{\rm eq}  \big|  E_k,\dots,E_{k+L-1},   \pi_k\not\in\Pi_{\rm eq} \big] \geq p_{\min}.\label{eq:ne2e}
\end{equation}
Fix $\tilde\epsilon\in(0,\epsilon)$ satisfying $\Big[\frac{(1-\tilde{\epsilon})  p_{\min}}{\tilde{\epsilon}+(1-\tilde{\epsilon})  p_{\min}}-\tilde{\epsilon}\Big][1-\tilde{\epsilon}] \geq 1-\epsilon$ , and choose  $\tilde{T}$ as above with the assumption $\min_\ell T_\ell\geq\tilde{T}$. We then obtain that for all $k$,
\begin{align}
& P\left[  \pi_{k+L} \in \Pi_{\rm eq}   |  \pi_k\not\in\Pi_{\rm eq}  \right] \geq p_{\min}(1-\tilde{\epsilon}) \nonumber \\
& \mbox{and} \ P\left[  \pi_{k+L} = \cdots = \pi_k  |  \pi_k\in\Pi_{\rm eq}  \right] \geq 1-\tilde{\epsilon}, \label{eq:neeqx}
\end{align}
Let $p_k:=P\left[  \pi_{k} \in \Pi_{\rm eq}  \right]$. With some manipulation, one can show:
\begin{enumerate}
\item $p_{(n+1)L}\geq p_{nL}+p_{\min}\tilde{\epsilon}$ as long as $p_{nL}<\frac{(1-\tilde{\epsilon})  p_{\min}}{\tilde{\epsilon}+(1-\tilde{\epsilon})  p_{\min}}$
\item $p_{(n+1)L}\geq p_{nL}-\tilde\epsilon$ as long as $p_{nL}\geq \frac{(1-\tilde{\epsilon})  p_{\min}}{\tilde{\epsilon}+(1-\tilde{\epsilon})  p_{\min}}$
\end{enumerate}
Thus there exists some $\tilde{n}<\infty$ such that for all $n\geq\tilde{n}$,
\[p_{nL}\geq \frac{(1-\tilde{\epsilon})  p_{\min}}{\tilde{\epsilon}+(1-\tilde{\epsilon})  p_{\min}}-\tilde{\epsilon}\]
and further that for all $n\geq\tilde{n}$, and $\ell\in\{1,\dots,L-1\}$
\[p_{nL+\ell}\geq \left(\frac{(1-\tilde{\epsilon})  p_{\min}}{\tilde{\epsilon}+(1-\tilde{\epsilon})  p_{\min}}-\tilde{\epsilon}\right)(1-\tilde{\epsilon})\geq 1-\epsilon.\]

{\it Part (ii) \citet[Appendix B]{ArslanYukselTAC16}} For any $\epsilon>0$, let $\tilde{T}<\infty$, $\tilde{k}<\infty$ be as in part (i). Let $\hat{k}<\infty$ be such that $\min_{k\geq\hat{k}}T_k\geq\tilde{T}$. It is straightforward to see from the proof of part (i) that for all $k\geq\hat{k}+\tilde{k}$, we have $P\left[ \pi_{k} \in \Pi_{\rm eq} \right] \geq 1 -\epsilon.$

{\it Part (iii) \citet[Appendix B]{ArslanYukselTAC16}}
Pick a sequence of $\{\tilde{\epsilon}_n\}_{n\geq0}\subset\mathbb{R}_{>0}$ satisfying  \begin{align*}
\sum_{n} (1-p_{\min})^{-n} \tilde{\epsilon}_n < \infty
\end{align*}
where $p_{\min}$ is as in (\ref{eq:ne2e}). As in {\it Part (i)}, there exist finite integers $\{\tilde{T}_k\}_{k\geq0}$ such that
\begin{align*}
& T_{nL},\dots,T_{(n+1)L-1}\geq \tilde{T}_n \\
& \Rightarrow \quad  P\left[  E_{nL},\dots,E_{(n+1)L-1}  \right]  \geq 1-\tilde{\epsilon}_n, \forall n
\end{align*}
This allows one to upper bound
\begin{align*}
& \sum_{k\geq1} P\left[  \pi_k \notin \Pi_{eq}  \right]  <\infty\quad \text{and}\quad \sum_{k\geq0} P\left[ \Omega\backslash E_k\right] <\infty.
\end{align*}
Borel-Cantelli Lemma implies the events $ \pi_k \notin \Pi_{eq}$ and $\Omega~\setminus~E_k$ each only occur for finitely many $k$, giving the desired result.\hfill$\diamond $

\section{Simulation}
We consider a modified version of the discounted stochastic game with two DMs as defined in Example~\ref{example:stochasticGenwag} with $\mathbb{U}^1=\mathbb{U^2}=\{1,2,3\}$, and $\mathbb{X}=\{1,2\}$, and discount factors $\beta^1=\beta^2=0.5$.

The state evolves as
\begin{align*}
	&P[x_{t+1}=1|x_t=1,(u_t^1,u_t^2)] = (1-\gamma)\mathbf{1}_{\{u_t^1=1 \text{ or } u_t^2=1\}}+\gamma/2\\
	&P[x_{t+1}=2|x_t=2,(u_t^1,u_t^2)] = (1-\gamma)\mathbf{1}_{\{(u_t^1,u_t^2)\neq(1,1)\}}+\gamma/2
\end{align*}
where $\gamma\in[0,1)$ introduces stochasticity, and $P[x_0]=[a,1-a]$ with $a$ chosen uniformly on $[0,1]$ at the start of the simulation. As such,  
the single-stage cost for each DM is given as 
\begin{figure}[h]
	\centering
	\begin{subfigure}[b]{0.3\textwidth}
		\centering
		{\small
			\begin{game}{3}{3}[$x=1$]
				& $1$      & $2$      & $3$ \\
				$1$     & 0, 0      & 9, 9       & 9, 9 \\
				$2$     & 9, 9      & 8, 9       & 9, 8 \\
				$3$     & 9, 9      & 9, 8       & 8, 9 \\
			\end{game}
		}
	\end{subfigure}
	\begin{subfigure}[b]{0.3\textwidth}
		\centering
		{\small
			\begin{game}{2}{2}[$x=2$]
				& $1$       & $2,3$       \\
				$1$     & 9, 9      & 8, 8     \\
				$2,3$     & 8, 8      & 9, 9      \\
			\end{game}
		}
	\end{subfigure}
	\caption{The single-stage cost of each DM, where DM$^1$ chooses the row, and DM$^2$ chooses the column.} \label{fig:simulationnon-acyclic-markov-game}
\end{figure}
For $x_0=1$, the single stage game corresponds to the discoordination game in Figure\ref{fig:non-trivial-game}, and for $x_0=2$, it is an anti-coordination game where players wish to choose the opposite actions. In the multi-stage game with $\gamma=0$, agents that follow better-response dynamics have no guarantee of reaching equilibrium from an arbitrary initial policy.  For small $\gamma$, the game remains GenWAG while still failing to be weakly acyclic. With this setup, there exists one Markov perfect equilibrium, at the joint policy where DM$^i$ chooses $u^i=1$ regardless of state.

We simulate Algorithm~ \ref{al:2GenWAGPure} with the following parameters: $\rho^i=0.1$, $\delta^i=10^{-8}$, $e^i=0.5$, , $\alpha^i_{k}=1/k^{0.751}$, for all $i,k$. We also keep the exploration phases of constant length for all $k$, with $T_k=T$, and reset Q-values to the the maximum cost at the end of each exploration phase.
Note that each DM has 9 possible policies $\pi^i:\mathbb{X}\rightarrow\mathbb{U}^i$,  and there are thus 81 possible joint strategies. We run the simulation for each choice of $\pi_0$, keeping track of whether $\pi_k$ is an equilibrium policy for each $k$, then take the average over all simulation runs. Figure~\ref{fig:simulation-results} demonstrates our results for $T=30000$, $k=300$, $\gamma=0.25$, revealing that equilibrium policies are reached more consistently for later exploration phases. 

\begin{figure}\label{fig:simulation-results}
	\centering
	\includegraphics[width=0.75\linewidth]{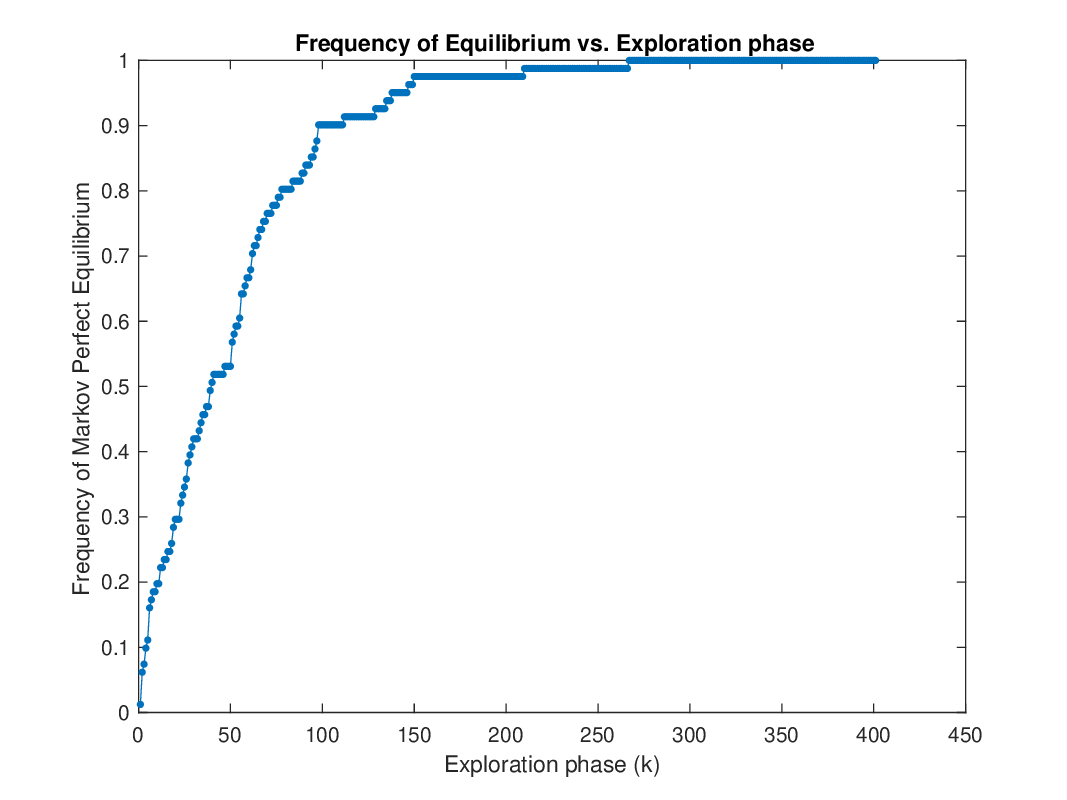}
	\caption{The fraction of times an equilibrium is reached at each exploration phase}
	\label{fig:simulation-results}
\end{figure}

\section{Beyond Deterministic Policies: $\epsilon$-Satisficing Paths to Equilibrium, and Implications for Learning}\label{sec:epsilon-paths}

In this paper, we have so far examined only the finite game $\Gamma$, and not the mixed extension where players are allowed to select their actions according to a probability distribution. In this section, we reflect on extensions to randomized policies and lay out several questions for future work. 

With a more general setup where players are allowed the choice of mixed strategies, it has been shown in \citet{yongacoglu2024paths} that any finite normal form game admits a satisficing path from any initial action profile to equilibrium. This result is established by demonstrating that at a strategy where the set of satisfied players cannot be made smaller via satisficing dynamics, there exists a sequence of strategies (where exactly these satisfied players remain satisfied) whose limit is an equilibrium strategy. 

It remains however to be shown when this property holds for $\epsilon$-satisficing paths, wherein players remain at the same strategy if they are $\epsilon$-best responding. For a maximizing agent, an action $a^i_{\star}$ is called an $\epsilon$-\emph{best response} to $\ba^{-i}$  when $r^i ( a^i_{\star}, \ba^{-i} ) \geq r^i ( \bar{a}^i, \ba^{-i} )-\epsilon$ for every $\bar{a}^i \in \aa^i$.

For pure strategies satisficing paths are robust to small perturbations and therefore an $\epsilon$-satisficing graph is the same as the $0$-satisficing graph for sufficiently small $\epsilon > 0$. In the mixed extension of $\Gamma$, players are allowed to choose from uncountably many policies, chosen from the set of probability distributions over their finite action sets. When applying this to learning, difficulties arise due to the uncountable nature of the mixed space, as each policy is assigned measure zero when randomizing over the policy space. Establishing conditions under which games have this $\epsilon$-satisficing paths property allows one to consider approximations that do not have this issue. 

It has been shown in \citet{yongacoglu2023satisficing} that all two-player games and all symmetric $n$-player games possess the $\epsilon$-satisficing paths property over mixed strategies. Additionally, \citet{yongacoglu2023satisficing} shows that when players are restricted to sufficiently fine quantizations, symmetric games retain the $\epsilon$-satisficing paths property. This allows players to learn over finite subsets of their policy spaces, and via policy revision. However, there are various difficulties in extending such results to general $n$-player games, with the proof used in \citet{yongacoglu2023satisficing} relying on a symmetry condition that is specific to symmetric games. The proof used in \citet{yongacoglu2024paths} also discusses the difficulties in establishing the $\epsilon$-satisficing paths property for general $n$-player games, as their proof for $0$-satisficing paths utilizes an indifference condition that cannot be generalized for $\epsilon >0$. Similarly, the sufficient condition for pure strategy games given in Theorem~\ref{theorem:n-player-weaker-condition} does not have a clear or useful extension for $\epsilon>0$.  It is currently an open question whether general results for $\epsilon$-satisficing paths can be achieved under either mixed or pure strategies, due to these challenges in extending the existing proof methods.

In the following, we discuss the learning theoretic implications of such a study for a class of games which are $\epsilon$-GenWAG.

\subsection{GenWAGs for General Stochastic Games and Convergence to Near Equilibria via Satisficing Paths} We now generalize the model studied in the previous section. As in the previous section denote a finite stochastic game with discounted cost by \vspace{-7.5pt}   		
\begin{equation}  \label{def:stochasticGame}
	\GG = ([n], \xx, \{ \mathbb{U}^i , c^i , \beta^i  : i \in [n] \}, P, \nu_0),
	\vspace{-5pt}
\end{equation}
where $[n]$ is a finite set of $n \in \nn$ DMs or agents, $\xx$ is a finite set of states; for DM$i$, $\uu^i$ is a finite set of actions, and we write $\bU := \times_{i \in [n]} \uu^i$. An element $\bu \in \bU$ is called a {\it joint action}. For DM$i$, $c^i : \xx \times \bU \to \rr$ is a stage cost function, and $\beta^i \in [0,1)$ is a discount factor. A random initial state $x_0 \in \xx$ is given by $x_0 \sim \nu_0$ where $\nu_0 \in \PP(\xx)$. State transitions are governed by the transition kernel $P \in \PP ( \xx | \xx \times \bU )$.

In the finite stochastic game above, each DM$i \in [n]$ uses a {\it policy} to select its sequence of actions $\{ u^i_t\}_{t \geq 0}$, using only information that is locally available at the time of each decision. We use $y^i_t$ to denote the observation variable for DM$i$ at time $t \geq 0$, and we let $h^i_t$ denote DM$i$'s information variable at time $t$, according to which DM$i$ selects $u^i_t$ under a policy.  

\begin{assumption}\label{ass:independent-learners}
	For each $i \in [n]$, DM$i$'s observation variables $\{ y^i_t \}_{t \geq 0}$ and information variables $\{h^i_t\}_{t \geq 0}$ are given by 
	\begin{description}
		\item[(i)] $y^i_0 = x_0$ and {\it $y^i_{t+1} = ( u^i_t, c^i ( x_t, \bu_t ), x_{t+1})$}\\ for $t \geq 0$;
		\item[(ii)] $h^i_0 = y^i_0$ and $h^i_{t+1} = ( h^i_t, y^i_{t+1})$ for $t \geq 0$;
	\end{description}
\end{assumption}

In the following, let $\Gamma^i_{S}$ denote stationary policies for DM$i$ and let $\bm{\Gamma}_s$ denote the collection of stationary policies of all agents. 

\begin{definition}
	Let $i \in [n]$, $\epsilon \geq 0$, and let $\Pi^i \subseteq \Gamma^i_{S}$. For $\bm{\pi}^{-i} \in \bm{\Gamma}^{-i}_{S}$, a policy $\tilde\pi^{i}\in \Pi^i$ is called an $\epsilon$-{\it best-response} for DM$i$ to $\bm{\pi}^{-i}$ {\it over $\Pi^i$} if 
	\begin{equation} 
		J^i  ( \tilde\pi^{i}, \bm{\pi}^{-i}  ,  x )  \leq \inf_{\pi^i\in \Pi^i} J^i (\pi^{i},\bm{\pi}^{-i} ,  x )  + \epsilon, \quad \forall x \in \xx.
	\end{equation}
\end{definition}

\begin{definition}
	For fixed $i \in [n]$, $\epsilon \geq0$, $\Pi^i \subseteq \Gamma^i_{S}$, and $\bm{\pi}^{-i} \in \bm{\Gamma}^{-i}_{S}$, we let $\best^i_{\epsilon} ( \bm{\pi}^{-i} , \Pi^i )$ denote DM$i$'s (possibly empty) set of $\epsilon$-best-responses to $\bm{\pi}^{-i}$ over $\Pi^i$. 
\end{definition}

\begin{definition}
	Let $\epsilon \geq 0$. A joint policy $\bm{\pi}_\star \in \bm{\Gamma}_{S}$ constitutes an $\epsilon$-equilibrium if $\pi^{i}_\star \in \best^i_{ \epsilon} ( \bm{\pi}^{-i} _\star, \Gamma^i_{S} )$ for every DM$i \in [n]$. 
\end{definition}

When $\epsilon = 0$, a $0$-best-response is simply called a best-response and a $0$-equilibrium is called an equilibrium. When the set $\Pi^i$ over which DM$i$ is optimizing is clear from context (typically, $\Pi^i = \Gamma^i_{S}$), we may omit ``over $\Pi^i$'' and simply write $\best^i_{\epsilon}  ( \bm{\pi}^{-i})$. 

\begin{definition}
	Let $\epsilon \geq 0$ and $T^i:\mathbf{\Pi}\rightarrow\Pi^i$ be a policy update rule for DM$i \in [n]$. The policy update rule $T^i$ is said to be {\it $\epsilon$-satisficing} if, for any $\bm{\pi} = ( \pi^i, \bm{\pi}^{-i} ) \in \bm{\Gamma}_{S}$, we have that $\pi^i \in \best^i_{\epsilon} ( \bm{\pi}^{-i} )$ implies $T^i ( \bm{\pi} ) = \pi^i.$
\end{definition}

A policy revision process is called $\epsilon$-satisficing if it is associated with a collection of policy update rules $\bT = \{ T^i \}_{i \in [n]}$ such that $T^i$ is $\epsilon$-satisficing for each DM$i \in [n]$. 

\begin{definition} \label{def:epsilon-path}
	Let $\epsilon \geq 0$. A (possibly finite) sequence $( \bm{\pi} )_{k \geq 0}$ of stationary joint policies is called an {\it $\epsilon$-satisficing path} if, for every $k \geq 0$ and $i \in \mathcal[n]$, $\pi^i_k \in \best^i_{\epsilon} ( \bm{\pi}^{-i}_k )$ implies $\pi^i_{k+1} = \pi^i_k$.
\end{definition}

\begin{definition} \label{def:paths-property}
	Let $\epsilon \geq 0$ and let $\bm{\Pi} \subseteq \bm{\Gamma}_{S}$ be a subset of stationary joint policies. A game $\GG$ is said to have the $\epsilon$-satisficing paths property within $\bm{\Pi}$ if for every $\bm{\pi} \in \bm{\Pi}$, there exists an $\epsilon$-satisficing path $( \bm{\pi}_t )_{t \geq 0}$ and an integer $K = K(\bm{\pi})$, such that (i)~$\bm{\pi}_0=\bm{\pi}$, (ii) $\bm{\pi}_t \in \bm{\Pi}$ for every $t \geq 0$, and (iii) $\bm{\pi}_K \in {\bf \Gamma}^{\epsilon-eq}_{S}$ (which is the set of $\epsilon$-equilibrium stationary policies).   
\end{definition}

In some applications, the environment being modelled exhibits symmetry in the agents. For such symmetric games: (1) each agent has the same set of actions; (2) the state dynamics depend only on the profile of actions taken by all players, without special dependence on the identities of the agents. That is, permuting the agents' actions in a joint action leaves the conditional probabilities for the next state unchanged; and (3) such a permutation results in a corresponding permutation of costs incurred. If $\uu^i = \uu^j$ for all $i,j \in [n]$, given a permutation $\sigma : [n] \to [n]$ and joint action $\ba = (a^i)_{i \in [n]}$, we define $\sigma( \ba) \in \bU$ to be the joint action in which DM$i$'s component is given by $ a^{\sigma(i)}$. That is, DM$i$'s action in $\sigma(\ba)$ is given by DM$\sigma(i)$'s action in $\ba$: $\sigma( \ba )^i = a^{\sigma(i)}$.

\begin{definition}[Symmetric Game] \label{def:symmetric game}
	A stochastic game $\GG$ defined by \eqref{def:stochasticGame} is called {\it symmetric} if the following holds:
	\begin{description} 
		\item[(i)] For any two players $i, j \in [n]$, we have $\beta^i = \beta^j$ and $\uu^i = \uu^j$; 
		
		\item[(ii)] For any $i \in [n]$, permutation $\sigma$ on $[n]$, and {\it $(\bs, \ba ) \in \xx \times \bU$}, we have $ c^i ( x, \sigma( {\it \ba}) ) = c^{\sigma(i)} ( x, {\it \ba} ) $ and $P \left( \cdot | x, {\it \ba}   \right) = P \left( \cdot | x , \sigma({\it \ba} )  \right).$
		
	\end{description}
\end{definition}

\begin{lemma} \label{lemma:symmetric-equality}
	Let $\GG$ be a symmetric game and $\bm{\pi} \in \bm{\Gamma}_{S}$ be a stationary joint policy. For $i, j \in [n]$, if $\pi^i  = \pi^j$, then $J^i ( \pi^i, \bm{\pi}^{-i}  ,  x ) = J^j (\pi^j, \bm{\pi}^{-j}  , x )$, for any $x \in \xx.$
\end{lemma}

\begin{corollary} \label{corollary:symmetric-equality} 
	Let $\GG$ be a symmetric game and let $\bm{\pi} \in \bm{\Gamma}_{S}$ be a joint policy. For $i, j \in [n]$, if $\pi^i  = \pi^j$, then $ \pi^i \in \best^i_{\epsilon} ( \bm{\pi}^{-i}, \Gamma^i_{S} )  \iff \pi^j \in \best^j_{\epsilon} ( \bm{\pi}^{-j}, \Gamma^j_{S} ) .$
\end{corollary}

We have two classes of games which exhibit the $\epsilon$-satisficing paths property: symmetric games and games with two players:

\begin{theorem} \label{thm:symmetric-paths} \citet{yongacoglu2021satisficing}
	Let $\GG$ be a symmetric stochastic game defined by \eqref{def:stochasticGame}. Then $\GG$ has the $\epsilon$-satisficing paths property in $\bm{\Gamma}_{S}$ for all $\epsilon \geq 0$. 
\end{theorem}

\begin{theorem} \label{thm:generalized-symmetric-paths}\citet{yongacoglu2021satisficing}
	Let $\GG$ be a symmetric game defined by \eqref{def:stochasticGame}, and let $\epsilon \geq 0$. Let $\bm{\Pi} \subseteq \bm{\Gamma}_{S}$ be a subset of stationary joint policies satisfying $\Pi^i = \Pi^j$ for all $i, j \in [n]$ and suppose $\bm{\Pi} \cap {\bf \Gamma}^{\epsilon-eq}
	_{S} \not= \varnothing$. Then, $\GG$ has the $\epsilon$-satisficing paths property within $\bm{\Pi}$. 
\end{theorem}

\begin{theorem}  \label{thm:two-player-paths}\citet{yongacoglu2021satisficing}
	Let $\GG$ be a stochastic game defined by \eqref{def:stochasticGame} with $n = 2$. Then, $\GG$ has the $\epsilon$-satisficing paths property within $\bm{\Gamma}_{S}$ for any $\epsilon \geq 0$.
\end{theorem}

One can design decentralized learners that exploit the $\epsilon$-satisficing paths property, which we show in the following. 

\subsection{Quantization of Stationary Policies}
The set of stationary policies is uncountable, but it could be made countable by quantization using a mapping $q^i : \Gamma^i_{S} \to \Gamma^i_{S}$. In the following, let $d^i$ be an appropriate metric on the spaces of policies $\Gamma^{i}_{S}$ for which the expected cost is continuous on $\Gamma^{i}_{S}$ under this metric.

\begin{definition}
	Let $\xi > 0$, $i \in [n]$, and $\tilde{\Pi}^i \subseteq \Gamma^i_{S}$. A mapping $q^i : \tilde{\Pi}^i \to \Gamma^{i}_{S}$ is called a {\it $\xi$-quantizer} (on $\tilde{\Pi}^i$) if 
	\begin{description}
		\item[(i)] the set $q^i ( \Gamma^i_{S} ) = \{ q^i ( \pi^i ) : \pi^i \in \Gamma^{i}_{S} \}$ is finite, and 
		\item[(ii)] for all $\pi^i \in \Gamma^i_{S}$, we have that $d^i ( \pi^i, q^i ( \pi^i )) < \xi$.  
	\end{description}
\end{definition}

\begin{definition} 
	Let $i \in [n]$, $\xi > 0$, and let $\tilde{\Pi}^i \subseteq \Gamma^i_{S}$. A set of policies $\Pi^i \subseteq \Gamma^i_{S}$ is called a {\it $\xi$-quantization} of $\tilde{\Pi}^i$ if $\Pi^i = q^i ( \tilde{\Pi}^i  )$ for some $\xi$-quantizer $q^i$.
\end{definition}

For any $\epsilon > 0$, there exists $\xi > 0$ such that if $\bm{\Pi} \subset \bm{\Gamma}_{S}$ is a $\xi$-quantization of $\bm{\Gamma}_{S}$, then $\bm{\Pi} \cap {\bf \Gamma}^{\epsilon-eq}_{S} \not= \varnothing$. Thus, if the quantization is fine enough, the game contains an $\epsilon$-equilibrium in the quantized policy space. Furthermore, let $\mathcal{G}$ be a symmetric stochastic game. If $\bm{\Pi}$ is a sufficiently fine quantization of $\bm{\Gamma}_{S}$ and $\Pi^j = \Pi^i$ for all $i, j$, then $\mathcal{G}$ has the $\epsilon$-satisficing paths property within $\bm{\Pi}$. That is, quantization and satisficing are compatible for symmetric games. This is useful for algorithm design, when the finiteness of the policy set can be exploited. In particular, Algorithm~\ref{al:2GenWAGPure} can applied for such problems leading to an equilibrium (which, however, may be subjective) \citet{yongacoglu2021satisficing}. 


\section{Conclusion}

We have presented GenWAGs, a generalization of the class of weakly acyclic games. We have demonstrated by example that this generalization is non-trivial, in the sense that there are examples of games that are GenWAGs but not weakly acyclic, and furthermore the class of GenWAGs does not include all games with pure Nash equilibrium. We showed that the class of GenWAGs is practically relevant for exploratory multi-agent learning applications because of its connection to the satisficing Markov chain, and we have provided multiple sufficient conditions for a game to be generalized weakly acyclic. We then presented positive implications for learning. 

	\vskip 0.2in
	\bibliography{GenWAGs,referencesGurdal}

\end{document}